\documentclass[12pt]{elsarticle}   

\usepackage[margin=3cm]{geometry}

\usepackage{array}
\usepackage{booktabs}
\usepackage{lineno,hyperref}
\modulolinenumbers[5]
\usepackage{mathtools}

\usepackage{graphicx}
\usepackage{caption}
\usepackage{xcolor}
\usepackage{subcaption}

\usepackage{natbib}

\usepackage{algorithm}
\usepackage{algcompatible}

\algnewcommand\INPUT{\item[\textbf{Input:}]}%
\algnewcommand\OUTPUT{\item[\textbf{Output:}]}%

\usepackage{graphicx}
\usepackage{amssymb}
\usepackage{bbm}
\usepackage{amsmath}

\newcommand{\Prt}[1]{\left({#1}\right)}

\newcommand{\Brc}[1]{{\left\{{#1}\right\}}}

\newcommand{\pd}{\partial}
\newcommand{\beql}[1]{\begin{equation}\label{#1}}
\newcommand{\eeq}{\end{equation}}
\newcommand{\beqr}[2]
{\begin{equation}\label{#1}\begin{array}{#2}}
\newcommand{\eeqr}
{\end{array}\end{equation}}
\newcommand{\bry}[1]
{\begin{array}{#1}}
\newcommand{\ery}{\end{array}}

\newcommand{\bX}{{\bf X}}

\newcommand{\tn}{t_n}
\newcommand{\pt}{t_{n-1}}

\newcommand{\tint}{t\in(\pt,\tn)}

\begin{document}

\begin{frontmatter}

\title{A note on the impact of management fees on the pricing of variable annuity guarantees}

\author[utsaddr,csiroaddr]{Jin Sun\corref{corrauth}}
\ead{jin.sun@uts.edu.au}
\author[macaddr]{Pavel V. Shevchenko}
\cortext[corrauth]{Corresponding author}
\ead{Pavel.Shevchenko@mq.edu.au}
\author[csiroaddr]{Man Chung Fung}
\ead{Simon.Fung@data61.csiro.au}

\address[utsaddr]{Faculty of Sciences, University of Technology Sydney, Australia}
\address[csiroaddr]{Data61, CSIRO, Australia}
\address[macaddr]{Department of Applied Finance and Actuarial Studies, Macquarie University, Australia}

\begin{abstract}
Variable annuities, as a class of retirement income products, allow equity market exposure for a policyholder's retirement fund with electable additional guarantees to limit the downside risk of the market. Management fees and guarantee insurance fees are charged respectively for the market exposure and for the protection from the downside risk. We investigate the impact of management fees on the pricing of variable annuity guarantees under optimal withdrawal strategies. Two optimal strategies, from policyholder's and from insurer's perspectives, are respectively formulated and the corresponding pricing problems are solved using dynamic programming. Our results show that when management fees are present, the two strategies can deviate significantly from each other, leading to a substantial difference of the guarantee insurance fees. This provides a possible explanation of lower guarantee insurance fees observed in the market. Numerical experiments are conducted to illustrate our results.
\end{abstract}


\end{frontmatter}

\nolinenumbers

\section{Introduction}
\label{intro}
Variable annuities (VA) with guarantees of living and death benefits are offered by wealth management and insurance companies worldwide to assist individuals in managing their pre-retirement and post-retirement financial plans. These products take
advantages of market growth while provide a protection of the savings against market
downturns. Similar guarantees are also available for life insurance policies (\citet{BacinelloOr96}). The VA contract cash flows received by the policyholder are linked to the investment portfolio choice and
performance (e.g. the choice of mutual fund and its strategy) while traditional annuities provide
a pre-defined income stream in exchange for a lump sum payment. Holders of VA policies are required to pay management fees regularly during the term of the contract for the wealth management services.

A variety of VA guarantees, also known as VA riders, can be elected by policyholders at the cost of additional insurance fees. Common examples of VA guarantees include guaranteed minimum accumulation benefit (GMAB), guaranteed minimum withdrawal benefit (GMWB), guaranteed minimum income benefit (GMIB) and guaranteed minimum death benefit (GMDB), as well as a combination of them. These guarantees, generically denoted as GMxB, provide different
types of protection against market downturns, shortfall of savings due to longevity risk or assurance of stability of income streams. Precise specifications of these products can vary across categories and issuers. See \citet{bauerKR2008,Ledlie2008,Kalberer2009} for an overview of these products.

The Global Financial Crisis during 2007-08 led to lasting adverse market conditions such as low interest rates and asset returns as well as high volatilities for VA providers. Under these conditions, the VA guarantees become more valuable, and the fulfillment of the corresponding required liabilities become more demanding. The post-crisis market conditions have called for effective hedging of risks associated with the VA guarantees (\citet{Sunetal16}). As a consequence, the need for accurate estimation of hedging costs of VA guarantees has become increasingly important. Such estimations consist of risk-neutral pricing of future cash flows that must be paid by the insurer to the policyholder in order to fulfill the liabilities of the VA guarantees.

There have been a number of contributions in the academic literature considering the pricing of VA guarantees. A range of numerical methods are considered, including standard and regression-based Monte Carlo (\citet{HuangKw16}), partial differential equation (PDE) and direct integration methods (\citet{MilSal2006,DaiKwok2008,ChenForsyth2008,bauerKR2008,LuoShev2015a,LuoShev2015b,ForsythV2014,LuoShev2016}). A comprehensive overview of numerical methods for the pricing of VA guarantees is provided in \citet{ShevLuo2016}.

In this article we focus on GMWB which provides a guaranteed withdrawal amount per year until the maturity of the contract regardless of the investment performance. The guaranteed withdrawal amount is determined such that the initial investment is returned over the life of the contract. When pricing GMWB, one typically assume either a pre-determined (static) policyholder behavior in withdrawal and surrender, or an active (dynamic) strategy where the policyholder ``optimally'' decides the amount of withdrawal at each withdrawal date depending on the information available at that date.

One of the most debated aspects in the pricing of GMWB with active withdrawal strategies is the policyholders' withdrawal behaviors (\citet{CramerMR2007,ChenForsyth2008,MoenigB2015,ForsythV2014}). It is often customary to refer to the withdrawal strategy that maximizes the expected liability, or the hedging cost, of the VA guarantee as the ``optimal'' strategy. Even though such a strategy underlies the worst case scenario for the VA provider with the highest hedging cost, it may not coincide with the real-world behavior of the policyholder. Nevertheless, the price of the guarantee under this strategy provides an upper bound of hedging cost from the insurer's perspective, which is often referred to as the ``value" of the guarantee. The real-world behaviors of policyholders often deviate from this ``optimal'' strategy, as is noted in \citet{MoenigB2015}. Different models have been proposed to account for the real-world behaviors of policyholders, including the reduced-form exercise rules of \citet{lee2005}, and the subjective risk neutral valuation approach taken by \citet{MoenigB2015}. In particular, it is concluded by \citet{MoenigB2015} that a subjective risk-neutral valuation methodology that takes different tax structures into consideration is in line with the corresponding findings from empirical observations.

Similar to the tax consideration in \citet{MoenigB2015}, the management fee is a form of market friction that would affect policyholders' rational behaviors. However, management fees are rarely considered in the VA pricing literature. When the management fee is zero and deterministic withdrawal behavior is assumed, \citet{HyndmanWe14} and \citet{FungIgSh14} show that risk-neutral pricing of guaranteed withdrawal benefits in both a policyholder's and an insurer's perspectives will result in the same fair insurance fee. Few studies that take management fees into account in the pricing of VA guarantees include \citet{BelangerFoLa09}, \citet{ChenVeFo08} and \citet{KlingRuRu11}. In these studies, fair insurance fees are considered from the insurer's perspective with the given management fees. The important question of how the management fees as a form of market friction will impact withdrawal behaviors of the policyholder, and hence the hedging cost for the insurer, is yet to be examined in a dynamic withdrawal setting. The main goal of the paper is to address this question.

The paper contributes to the literature in three aspects. First, we consider two pricing approaches based on the policyholder's and the insurer's perspective. In the literature it is most often the case that only an insurer's perspective is considered, which might result in mis-characterisation of the policyholder's withdrawal strategies. Second, we characterize the impact of management fees on the pricing of GWMB, and demonstrate that the two afore-mentioned pricing perspectives lead to different fair insurance fees due to the presence of management fees. In particular, the fair insurance fees from the policyholder's perspective is lower than those from the insurer's perspective. This provides a possible justification of lower insurance fees observed in the market.  Third, the sensitivity of the fair insurance fees to management fees under different market conditions and contract parameters are investigated and quantified through numerical examples.

The paper is organized as follows. In Section \ref{formulation} we present the contract details of the GMWB guarantee together with its pricing formulation under a stochastic optimal control framework. Section \ref{pvalue} derives the policyholder's value function under the risk-neutral pricing approach, followed by the insurer's net liability function in Section \ref{insliab}. In Section \ref{strategies} we compare the two withdrawal strategies that maximize the policyholder's value and the insurer's liability, respectively, and discuss the role of the management fees in their relations. Section \ref{num} demonstrates our approaches via numerical examples. Section \ref{conc} concludes with remarks and discussion.

\section{Formulation of the GMWB pricing problem}
\label{formulation}

We begin with the setup of the framework for the pricing of GMWB and describe the features of this type of guarantees. The problem is formulated under a general setting so that the resulting pricing formulation can be applied to different GMWB contract specifications. Besides the general setting, we also consider a very specific simple GMWB contract, which will be subsequently used  for illustration purposes in numerical experiments presented in Section \ref{num}.

The VA policyholder's retirement fund is usually invested in a managed wealth account that is exposed to financial market risks. A management fee is usually charged for this investment service. In addition, if GMWB is elected, extra insurance fees will be charged for the protection offered by the guarantee provider (insurer).  We assume the wealth account guaranteed by the GMWB is subject to continuously charged proportional management fees, paid to an independent wealth manager other than the insurer. This assumption implies that the management fees cannot be used to fund the hedging portfolio for the GMWB guarantee. The sole purpose of management fees is to compensate for the fund management services provided, and should
not be confused with the hedging cost of the guarantee. The cost of hedging, on the the hand, is paid by proportional insurance fees continuously charged to the wealth account. The fair insurance fee rate, or the fair fees in short, refers to the minimal insurance fee rate required to fund the hedging portfolio, or the replicating portfolio, so that the guarantee provider can eliminate the market risk associated with the selling of the guarantees.

We consider the situation where a policyholder purchases the GMWB rider in order to protect his wealth account that tracks an equity index $S(t)$ at time $t\in[0,T]$, where $0$ and $T$ correspond to the inception and expiry dates. The equity index account is modelled under the risk-neutral probability measure $\mathbb{Q}$ following the stochastic differential equation (SDE)
\beql{equity}
dS(t)=S(t)\Prt{r(t)dt+\sigma(t) dB(t)},
\quad t\in[0,T],
\eeq
where $r(t)$ is the risk-free short interest rate, $\sigma(t)$ is the volatility of the index, which are made time-dependent and can be stochastic, and $B(t)$ is a standard $\mathbb{Q}$-Brownian motion modelling the uncertainty of the index. Here, we follow standard practices in the literature of VA guarantee pricing by modelling under the risk-neutral probability measure $\mathbb{Q}$, which allows the pricing of stochastic cash flows as taking the risk-neutral expectation of the discounted cash flows. The risk-neutral probability measure $\mathbb{Q}$ exists if the underlying financial market satisfies certain ``no-arbitrage'' conditions. For details on risk-neutral pricing, see, e.g., \citet{DelbaenSc06} for an account under very general settings. 

The wealth account $W(t),t\in[0,T]$ over the lifetime of the GMWB contract is invested into the index $S$, subject to management fees charged by a wealth manager at the rate $\alpha_{\rm m}(t)$. An additional charge of insurance fees at rate $\alpha_{\rm ins}(t)$ for the GMWB rider is collected by the insurer to pay for the hedging cost of the guarantee. Both fees are deterministic, time-dependent and continuously charged. Discrete fees may be modelled similarly without any difficulty. The wealth account in turn evolves as
\beql{wsde}
dW(t)=W(t)\Prt{(r(t)-\alpha_{\rm tot}(t))dt+\sigma(t)dB(t)},\quad \eeq
for any $t\in[0,T]$ at which no withdrawal of wealth is made. Here, $\alpha_{\rm tot}(t)=\alpha_{\rm ins}(t)+\alpha_{\rm m}(t)$ is the total fee rate. The GMWB contract allows the policyholder to withdraw from a guarantee account $A(t),t\in[0,T]$ on a sequence of pre-determined contract event dates, $0=t_0<t_1<\dots<t_N=T$. The initial guarantee $A(0)$ usually matches the initial wealth $W(0)$. The guarantee account stays constant unless a withdrawal is made on one of the event dates, which changes the guarantee account balance. We assume that the GMWB contract will be taken over by the beneficiary if the policyholder dies before the maturity $T$, so that no early termination of the contract is possible, nor is there any death benefit included in the contract. Additional features such as early surrender and death benefits can be included straightforwardly but will not be considered in this article, in order to better illustrate the impact of management fees without unnecessary complexities.

To simplify notations, we denote by $\bX(t)$ the vector of state variables at $t$, given by
\beql{statev}
\bX(t)=(r(t),\sigma(t),S(t),W(t),A(t)),\quad t\in[0,T].
\eeq
We denote by $\text{E}^\mathbb{Q}_t[\cdot]$ the risk-neutral expectation conditional on the state variables at $t$, i.e., $\text{E}^\mathbb{Q}_t[\cdot]:=\text{E}^{\mathbb{Q}}[\cdot|\bX(t)]$. Here, we assume that all state variables follow Markov processes under the risk-neutral probability measure $\mathbb{Q}$, so that $\bX(t)$ contains all the information available at $t$. For completeness, we include the index value $S(t)$ in the vector of state variables which under the current model may seem redundant, due to the scale-invariance of the geometric Brownian motion type model (\ref{equity}). In general, however, $S(t)$ may determine the future dynamics of $S$ in a nonlinear fashion, as is the case under, e.g., the minimal market model described in \citet{PlatenHe06}.

On event dates $\tn,n=1,\dots,N-1$, the policyholder may choose to withdraw a nominal amount $\gamma_n\le A(\tn)$. The real cash flow received by the policyholder is denoted by $C_n(\gamma_n,\bX(\tn^-))$, where $t^-$ refers to the time ``just before'' $t$.
As a specific example, $C_n(\gamma_n,\bX(\tn^-))$ may be given by
\beql{cf}
C_n(\gamma_n,\bX(\tn^-))=\gamma_n-\beta\max(\gamma_n-G_n,0),
\eeq
where the contractual withdrawal $G_n$ is a pre-determined withdrawal amount specified in the GMWB contract, and $\beta$ is the penalty rate applied to the part of the withdrawal exceeding the contractual withdrawal $G_n$. The policyholder may decide the withdrawal amount $\gamma_n$ based on all current state variables, i.e.,
\beql{strategy}
\gamma_n=\Gamma(\tn,\bX(\tn^-)),
\eeq
where the mapping $\Gamma(\cdot,\cdot)$ is defined as a \emph{withdrawal strategy}. Given the assumed Markovian structure of the state variables $\bX$, the withdrawal strategy (\ref{strategy}) uses all current information.

Upon withdrawal, the guarantee account is changed by the amount $D_n(\gamma_n,\bX(\tn^-))$, that is,
\beql{ga}
A(\tn)=A(\tn^-)-D_n(\gamma_n,\bX(\tn^-)).
\eeq
For example, if $D_n(\gamma_n,\bX(\tn^-))=\gamma_n$, then
\beql{gas}
A(\tn)=A(\tn^-)-\gamma_n.
\eeq
The guarantee account stays nonnegative, that is, $\gamma_n$ can only be taken such that $D_n(\gamma_n,\bX(\tn^-))\le A(\tn^-)$.
The wealth account is reduced by the amount $\gamma_n$ upon withdrawal and remains nonnegative. That is,
\beql{wjump}
W(\tn)=\max(W(\tn^-)-\gamma_n,0),
\eeq
where $W(\tn^-)$ is the wealth account balance just before the withdrawal. It is assumed that $\gamma_0=0$, i.e., no withdrawals at the start of the contract.

The function of policyholder's remaining value at time $t$ is denoted by $V(t,\bX(t)),t\in[0,T]$, which corresponds to the risk-neutral value of all future cashflows to the policyholder at time $t$.
At maturity $t_N=T$, the policyholder obtains an liquidation cash flow $V(T^-,\bX(T^-))$. Both account balances becomes zero at $T$, that is, $W(T)=A(T)=0$. For example, the liquidation cash flow may be given by
\beql{Qterm}
V(T^-,\bX(T^-))=A(T^-)-\beta\max(A(T^-)-G_N,0)+\max(W(T^-)-A(T^-),0),
\eeq
which assumes a nominal withdrawal $\gamma_N=A(T^-)$ of remaining guarantee account balance subject to a penalty implied by (\ref{cf}) with contract amount $G_N$, plus the liquidation of the wealth account after this withdrawal.
Different contracts may define different liquidation cash flows. We emphasize here that the analysis presented in this article does not require this particular form (\ref{Qterm}). Note also that the liquidation cash flow either does not depend on $\gamma_N$ or the dependence is only formal, in that $\gamma_N$ is always chosen to maximize the liquidation cash flow.

\section{The Policyholder's Value Function}
\label{pvalue}
Having introduced the modeling framework and contract specifications, we can now calculate the policyholder's value function $V(t,\bX(t))$ as the risk-neutral expected value of policyholder's future cash flows at time $t\in[0,T]$. Valuing the future cash flows under the risk-neutral pricing approach assumes the policyholder's cash flow may be replicated by self-financing portfolios of the same initial value. Although the policyholder may not have the resources to carry out any hedging strategies in person, we assume the market is liquid enough to trade such strategies without frictions, and the policyholder has access to the market through independent agents. Thus the risk-neutral valuation of the policyholder's future cash flows can be regarded as the value of the remaining term of the VA contract from the poliyholer's perspective.

Following Section~\ref{formulation}, for a given withdrawal strategy $\Gamma$, the jump condition at the withdrawal time $\tn$ for the policyholder's value function $V(t,\bX(t))$ is given by
\beql{Qjump}
V(\tn^-,\bX(\tn^-))=C_{n}(\gamma_{n},\bX(\tn^-))+V(\tn,\bX(\tn)),
\eeq
i.e., the remaining policy value just before the withdrawal is the sum of the cash flow from the withdrawal and the remaining policy value immediately after the withdrawal. Here, the withdrawal $\gamma_n$ is given as (\ref{strategy}) by applying the strategy of choice.

The policy value at $\tint$ is given by the discounted expected future policy value under the risk-neutral probability measure as
\beql{Qrecur}
V(t,\bX(t))=\text{E}^\mathbb{Q}_t\left[e^{-\int_t^{\tn}r(s)ds}V(\tn^-,\bX(\tn^-))\right],
\eeq
where $e^{-\int_t^{\tn}r(s)ds}$ is the discount factor. The initial policy value, given by $V(0,\bX(0))$, can be calculated backward in time starting from the terminal condtion $V(T^-,\bX(T^-))$, using (\ref{Qjump}) and (\ref{Qrecur}), as described in Algorithm \ref{algo1}.

As an illustrative example, we assume $r(t)\equiv r$, $\sigma(t)\equiv\sigma$ and $\alpha_{\rm tot}(t)\equiv\alpha_{\rm tot}$ as constants, and thus $V(t,\bX(t))=V(t,W(t))$ for $t\in(\pt,\tn)$. We now derive the partial differential equation (PDE) satisfied by the value function $V$ through a hedging argument. 
We consider a delta hedging portfolio that, at time $\tint$, takes a long position of the value function $V$ and a short position of $\frac{W(t)\pd_WV(t,W(t))}{S(t)}$ shares of the index $S$. Here $\pd_WV(t,W(t)) \equiv \frac{\pd V(t,W)}{\pd W}|_{W=W(t)}$ is the partial derivative of $V(t,W)$ with respect to the second argument, evaluated at $W(t)$. Denoting the total value of this portfolio at $\tint$ as $\Pi^V(t)$, the value of the delta hedging portfolio is given by
\beql{QhedgeP}
\Pi^V(t)=V(t,W(t))-W(t)\pd_WV(t,W(t)).
\eeq
By Ito's formula and (\ref{equity}), the SDE for $\Pi^V$ can be obtained as
\beqr{Qhedge}{l}
d\Pi^V(t)=\Prt{\pd_tV(t,W(t))-\alpha_{\rm tot}W(t)\pd_WV(t,W(t))+\frac{1}{2}\sigma^2W(t)^2\pd_{WW}V(t,W(t))}dt, \\
\eeqr
for $\tint$.  Since the hedging portfolio $\Pi^V$ is locally riskless, it must grow at the risk-free rate $r$, that is $d\Pi^V(t)=r\Pi^V(t)dt$. This along with (\ref{QhedgeP}) implies that the PDE satisfied by the value function $V(t,W)$ is given by
\beql{Qpde}
\pd_tV-rV+(r-\alpha_{\rm tot})W\pd_WV+\frac{1}{2}\sigma^2W^2\pd_{WW}V=0,
\eeq
for $\tint$ and $n=1,\dots,N$. The boundary conditions at $\tn$ are specified by (\ref{Qterm}) and (\ref{Qjump}). The valuation formula (\ref{Qrecur}) or the PDE (\ref{Qpde}) may be solved recursively by following Algorithm \ref{algo1} to compute the initial policy value $V(0,\bX(0))$. It should be noted that (\ref{Qrecur}) is general, and does not depend on the simplifying assumptions made in the PDE derivation.
\begin{algorithm}
	\caption{Recursive computation of $V(0,\bX(0))$}
	\begin{algorithmic}[1]
		\STATE choose a withdrawal strategy $\Gamma$
		\STATE initialize $V(T^-,\bX(T^-))$, e.g., as (\ref{Qterm})
		\STATE set $n=N$
		\WHILE{$n>0$}
		\STATE compute $V(\pt,\bX(\pt))$ by solving (\ref{Qrecur}) or (\ref{Qpde}) with terminal condition $V(\tn^-,\bX(\tn^-))$
		\STATE compute the withdrawal amount $\gamma_{n-1}$ by applying the strategy $\Gamma$ as (\ref{strategy}) 
		\STATE compute $V(\pt^-,\bX(\pt^-))$ by applying jump condition (\ref{Qjump})
		\STATE $n=n-1$
		\ENDWHILE
	\end{algorithmic}
	\label{algo1}
\end{algorithm}

\section{The Insurer's Liability Function and the Fair Fee Rate}
\label{insliab}
The GMWB contract may be considered from the insurer's perspective by examining the insurer's liabilities, given by the risk-neutral value of the cash flows that must be paid by the insurer in order to fulfill the GMWB contract.

On any withdrawal date $\tn$, the actual cash flow received by the policyholder is given by (\ref{cf}). This cash flow is first paid out of the policyholder's real withdrawal from the wealth account, which is equal to $\min(W(\tn^-),\gamma_n)$, the smaller of the nominal withdrawal and the available wealth. If the wealth account has an insufficient balance, the rest must be paid by the insurer. If the real withdrawal exceeds the cash flow entitled to the policyholder, the insurer keeps the surplus. The payment made by the insurer at $\tn$ is thus given by
\beql{cn}
c_n(\gamma_n,\bX(\tn^-))=C_n(\gamma_n,\bX(\tn^-))-\min(W(\tn^-),\gamma_n).
\eeq
At any time $t$, we denote the \emph{net liability} function as $L(t,\bX(t))$, which refers to the risk-neutral value of all future payments made to the policyholder by the insurer, less the present value of all future insurance fee incomes.

The insurance fees, charged at the rate $\alpha_{\rm ins}(t),\,t\in[0,T]$, is called fair if the total fees exactly compensate for the insurer's total liability, such that the net liability is zero at time $t=0$. That is,
\beql{fairalpha}
L(0,\bX(0))=0.
\eeq
Note that $L(0,\bX(0))$ depends on $\alpha_{\rm ins}(t)$, which was made implicit for notational simplicity. If $\alpha_{\rm ins}(t)\equiv\alpha_{\rm ins}$ is a constant, its value can be found by solving (\ref{fairalpha}).  The fair insurance fees represent the hedging cost for the insurer to deliver the GMWB guarantee to the policyholder, which is often regarded as the value of the GMWB rider, at least from the insurer's perspective. We emphasize here that this value may not be equal to the added value of the GMWB rider to the policyholder's wealth account, as we will show in Section \ref{strategies}.

To compute $L(0,\bX(0))$ we first note that at maturity $T$, the terminal condition on $L$ is given by
\beql{Lterm}
L(T^-,\bX(T^-))=V(T^-,\bX(T^-))-W(T^-),
\eeq
i.e., the insurer must pay for any amount of the policyholder's final value not covered by the available wealth. Depending on the GMWB contract details, this amount can be negative, in which case the insurer gets paid. This happens, for example, if the liquidation cash flow is given by (\ref{Qterm}), and there are more penalties applied to the final (compulsory) withdrawal due to forced liquidation of a high final balance of the guarantee account. The jump condition on $L$ at $\tn$ is given by
\beql{Ljump}
L(\tn^-,\bX(\tn^-))=c_n(\gamma_n,\bX(\tn^-))+L(\tn,\bX(\tn)),
\eeq
i.e., upon withdrawal, the net liability is reduced by the amount paid out.

At $\tint$, the net liability function is given by the risk-neutral value of the remaining liabilities at $\tn$ before any benefit is paid, less any insurance fee incomes over the period $(t,\tn)$, discounted at the risk-free rate. Specifically, we have
\beqr{Lrecur}{c}
L(t,\bX(t))=\text{E}^\mathbb{Q}_t\Big[e^{-\int_t^{\tn}r(s)ds}L(\tn^-,W(\tn^-),A(\tn^-))\Big]\\
\qquad\qquad\quad-\text{E}^\mathbb{Q}_t\Big[\int_{t}^{\tn}e^{-\int_t^{s}r(u)du}\alpha_{\rm ins}(s)W(s)ds\Big].
\eeqr
Note that the net liability, viewed from time $t$ forward, is reduced by expecting to receive insurance fees over $(t, \tn)$. Since this reduction decreases with time, the net liability \emph{increases} with time.

To give an example, we again assume constant $r(t)\equiv r$, $\sigma(t)\equiv \sigma$, $\alpha_{\rm ins}(t)\equiv \alpha_{\rm ins}$, $\alpha_{\rm tot}(t)\equiv \alpha_{\rm tot}$. Under these simplifying assumptions we have $L(t,\bX(t))=L(t,W(t))$, for $\tint$. To derive the PDE satisfied by $L(t,W)$, consider a delta hedging portfolio that, at time $\tint$, consists of a long position in the net liability function $L$ and a short position of $\frac{W(t)\pd_WL(t,W(t))}{S(t)}$ shares of the index $S$. The value of the delta hedging portfolio, denoted as $\Pi^L(t)$,  is given by
\beql{LhedgeP}
\Pi^L(t)=L(t,W(t))-W(t)\pd_WL(t,W(t)).
\eeq
By Ito's formula and (\ref{equity}), we obtain the SDE for $\Pi^L$ as
\beqr{Lhedge}{l}
d\Pi^L(t)=\Prt{\pd_tL(t,W(t))-\alpha_{\rm tot}W(t)\pd_WL(t,W(t))
	+\frac{1}{2}\sigma^2W(t)^2\pd_{WW}L(t,W(t))}dt,\\
\eeqr
where $\tint$. Since the hedging portfolio $\Pi^L$ is locally riskless and must grow at the risk-free rate $r$, as well as increase with the insurance fee income at rate $\alpha_{\rm ins}W(t)$ (see remarks after (\ref{Lrecur})), we must also have $d\Pi^L(t)=\Prt{r\Pi^L(t)+\alpha_{\rm ins}W(t)}dt$. This along with (\ref{LhedgeP}) implies that the PDE satisfied by the value function $L(t,W)$ is given by
\beql{Lpde}
\pd_tL-\alpha_{\rm ins} W-rL
+(r-\alpha_{\rm tot})W\pd_WL
+\frac{1}{2}\sigma^2W^2\pd_{WW}L=0,
\eeq
for $\tint$. The initial net liability can thus be computed by recursively solving (\ref{Lrecur}) or (\ref{Lpde}) from terminal and jump conditions (\ref{Lterm}) and (\ref{Ljump}), as described in Algorithm \ref{algo2}.
\begin{algorithm}
	\caption{Recursive computation of $L(0,\bX(0))$}
	\begin{algorithmic}[1]
		\STATE choose a withdrawal strategy $\Gamma$
		\STATE initialize $L(T^-,\bX(T^-))$ as (\ref{Lterm})
		\STATE set $n=N$
		\WHILE{$n>0$}
		\STATE compute $L(\pt,\bX(\pt))$ by solving (\ref{Lrecur}) or (\ref{Lpde}) with terminal condition $L(\tn^-,\bX(\tn^-))$
		\STATE compute the withdrawal amount $\gamma_{n-1}$ by applying the strategy $\Gamma$ as (\ref{strategy}) 
		\STATE compute $L(\pt^-,\bX(\pt^-))$ by applying jump condition (\ref{Ljump})
		\STATE $n=n-1$
		\ENDWHILE
	\end{algorithmic}
	\label{algo2}
\end{algorithm}

\section{Policy Value Maximization vs Liability Maximization}
\label{strategies}
In the previous sections, the withdrawal strategy $\gamma_n,n=1,\dots,n-1$ has been assumed to be given. The withdrawal strategy serves as a control sequence affecting the policyholder's value function and the insurer's liability function. These withdrawals may thus be chosen to maximize either of these functions, leading to two distinct withdrawal strategies. In this section we formulate these two strategies and discuss their relations. In particular, we point out the role of  management fees in the relations between the strategies and the implications. 

\subsection{Formulation of two optimization problems}
We first formulate the policyholder's value maximization problem, i.e., maximizing the initial policy value $V(0,\bX(0))$ by optimally choosing the sequence $\gamma_n$ for $n=1,\dots,N-1$.
Following the principle of dynamic programming, this is accomplished by choosing the withdrawal $\gamma_n$ as
\beql{ratgamma}
\gamma_n=\Gamma^V(\tn,\bX(\tn^-))=
\underset{\gamma\in\mathcal{A}}{\arg\max}\Brc{C_n(\gamma,\bX(\tn^-))+V\Prt{\tn,\bX(\tn|\bX(\tn^-),\gamma)}}
\eeq
in the admissible set  $\mathcal{A}=\Brc{\gamma:\gamma\ge0,A(\tn|\bX(\tn^-),\gamma)\ge0}$. Here, we used $\bX(\tn|\bX(\tn^-),\gamma)$ and $A(\tn|\bX(\tn^-),\gamma)$  to denote the state variables $\bX(\tn)$ and $A(\tn)$ at $\tn$ after withdrawal $\gamma$ is made, given the value of the state variables $\bX(\tn^-)$ before the withdrawal.
At any withdrawal time $\tn$ the policyholder chooses the withdrawal $\gamma\in\mathcal{A}$ to maximize the sum of the cash flow he receives and the present value of the remaining term of the policy. The strategy $\Gamma^V$ given by (\ref{ratgamma}) is called the \emph{value maximization strategy}.

On the other hand, the optimization problem from the insurer's perspective considers the most unfavourable situation for the insurer. That is, by making suitable choices of $\gamma_n$'s, the policyholder attempts to maximize the net initial liability function $L(0,\bX(0))$. Even though a policyholder has little reason to pursue such a strategy, the fair fee rate under this strategy is guaranteed to cover the hedging cost of the GMWB rider regardless of the withdrawal strategy of the policyholder (assuming the insurer can perfectly hedge the market risks). The withdrawal $\gamma_n$ for this strategy is given by
\beql{advgamma}
\gamma_n=
\Gamma^L(\tn,\bX(\tn^-))=
\underset{\gamma\in\mathcal{A}}{\arg\max}\Brc{c_n(\gamma,\bX(\tn^-))+L\Prt{\tn,\bX(\tn|\bX(\tn^-),\gamma)}},
\eeq
i.e., the sum of the cash flow paid by the insurer and the net liability of the remaining term of the contract is maximized. The strategy $\Gamma^L$ given by (\ref{advgamma}) is called the \emph{liability maximization strategy}.

\subsection{The role of management fees}
To clarify the different implications on the fair insurance fees between the value and the liability maximization strategies, we now establish the relationship between the policy value $V$ and the net liability $L$ by defining the process
\beql{pvm}
M(t,\bX(t)):=L(t,\bX(t))+W(t)-V(t,\bX(t)),
\eeq
for $t\in[0,T]$. From (\ref{Lterm}) we obtain
\beql{Mterm}
M(T^-,\bX(T^-))=0,
\eeq
as the terminal condition for $M$. From (\ref{Qrecur}) and (\ref{Lrecur}) we find the recursive relation for $M$ as,
\beqr{Mrecur0}{l}
M(t,\bX(t))=\text{E}^\mathbb{Q}_t\left[e^{-\int_t^{\tn}r(s)ds}M(\tn^-,\bX(\tn^-))\right]\\
\qquad\qquad\qquad+W(t)-\text{E}^\mathbb{Q}_t\left[W(\tn^-)\right]\\
\qquad\qquad\qquad-\text{E}^\mathbb{Q}_t\left[\int_{t}^{\tn}e^{-\int_t^{s}r(u)du}\alpha_{\rm ins}(s)W(s)ds\right].
\eeqr
Note that the second and third lines in (\ref{Mrecur0}) can be identified with the time-$t$ risk-neutral value of management fees over $(t,\tn)$. To see this, we first note that the difference of the first two terms is the time-$t$ risk-neutral value of the total fees charged on the wealth account over $(t,\tn)$, and the expectation in the third term is the time-$t$ risk-neutral value of the insurance fees over the same period.

In lights of (\ref{Mterm}) and (\ref{Mrecur0}), the quantity $M(t,\bX(t))$ defined by (\ref{pvm}) is precisely the time-$t$ risk-neutral value of future management fees. From (\ref{pvm}), the policy value may be written as
\beql{QV}
V(t,\bX(t))=W(t)+L(t,\bX(t))-M(t,\bX(t)),
\eeq
i.e., the sum of the wealth and the value of the GMWB rider, less the value of future management fees. At $t=0$, this gives
\beql{QV0}
V(0,\bX(0))+M(0,\bX(0))=W(0)+L(0,\bX(0)).
\eeq
Therefore, maximizing $L(0,\bX(0))$ in general is not the same as maximizing $V(0,\bX(0))$, since the total management fee $M(0,\bX(0))$ depends on the withdrawal strategy.
The fair fee condition (\ref{fairalpha}) becomes
\beql{fairalphaQ}
V(0,\bX(0))+M(0,\bX(0))=W(0),
\eeq
as in contrast to the $V(0)=W(0)$ condition often seen in the literature,  when no management fees are charged, in which case the two strategies $\Gamma^V$ and $\Gamma^L$ coincide. When the management fee rate is positive, the liability maximization strategy $\Gamma^L$ by definition leads to the maximal initial net liability, thus by (\ref{fairalpha}) the maximal fair insurance fee rate.

\subsection{Implications}
We now consider the two strategies in an idealized world where the policy value and the net liability processes can be perfectly replicated using self-financing trading strategies. As is mentioned above, assuming that the fair insurance fee follows from the liability maximization strategy, the insurer is guaranteed a nonnegative profit, regardless of the actual withdrawal strategies of the policyholder. If the policyholder behaves differently from this strategy, in particular, if he follows the value maximization strategy, the insurer generally makes a positive profit. 

On the other hand, consider the situation where the policyholder purchases his policy from a middle agent, who in turn purchases the same policy for the same price from an insurer in the name of the policyholder, and handles withdrawals at his own choice, but fulfills any withdrawal requests from the policyholder according to the GMWB contract. In other words, the middle agent provides the GMWB guarantee to the policyholder, and follows the value maximization strategy when making his own withdrawals from the insurer. Then regardless of the policyholder's withdrawal behavior or the fees, the middle agent always makes a nonnegative profit. If the policyholder behaves differently from the policy value maximization strategy, she receives a value less than the maximal policy value received by the agent, who therefore makes a positive profit out of the suboptimal behavior of the policyholder. Given that the middle agent will carry out withdrawals that would maximize the value rather than the cost to the insurer, the insurer in turn can afford to charge a less expensive fee than those implied by the liability maximization strategy, leading to more value for the middle agent and the policyholder. The seemingly win-win situation come at the loss of the wealth manager, who now expect to receive less management fees. In this case the middle agent in effect reduces the market frictions represented by the management fees by maximizing his own value, and at the same time help improving the policyholder's value.

\section{Numerical Examples}
\label{num}
To demonstrate the impact of management fees on the fair fees of GMWB contracts, we carry out in this section several numerical experiments. We investigate how the presence of management fees will lead to different fair fees for the two withdrawal strategies studied in previous sections under different market conditions and contract parameters.

\subsection{Setup of the experiments}
For illustration purposes, we assume a simple GMWB contract as specified by (\ref{cf}), (\ref{gas}), (\ref{Qterm}) as well as constant $r$, $\sigma$, $\alpha_{\rm m}$ and $\alpha_{\rm ins}$ so that the PDEs (\ref{Qpde}) and (\ref{Lpde}) are valid.
We consider different contractual scenarios and calculate the fair fees implied by (\ref{fairalpha}) under the withdrawal strategies given in Section \ref{strategies}.

It is assumed that the wealth and the guarantee accounts start at $W(0)=A(0)=1$. The maturities of the contracts range from 5 to 20 years, with annual contractual withdrawals evenly distributed over the lifetime of the contracts. The first withdrawal occurs at the end of the first year and the last at the maturity. The management fee rate ranges from $0\%$ up to $2\%$.

We consider several investment environments with the risk-free rate $r$ at levels $1\%$ and $5\%$, and the volatility of the index $\sigma$ at $10\%$ and $30\%$, to represent different market conditions such as low/high growth and low/high volatility scenarios. In addition, the penalty rate $\beta$ may take values at $10\%$ or $20\%$.

We compute the initial policy value $V(0,\bX(0))$ as well as the initial net liability $L(0,\bX(0))$ at time 0 numerically by following Algorithms \ref{algo1} and \ref{algo2} simultaneously. The withdrawal strategies $\Gamma^L$ and $\Gamma^V$ are considered separately. The PDEs (\ref{Qpde}) and (\ref{Lpde}) are solved using Crank-Nicholson finite difference method (\citet{crank1947,Hirsa}) with appropriate terminal and jump conditions for both functions under both strategies. This leads to the initial values and liabilities $V(0,\bX(0);\Gamma^L)$, $L(0,\bX(0);\Gamma^L)$,  $V(0,\bX(0);\Gamma^V)$ and $L(0,\bX(0);\Gamma^V)$ under both strategies. Here, we made the dependence of these functions on the strategies explicit.

The fair fee rates under both strategies were obtained by solving (\ref{fairalpha}) using a standard root-finding numerical scheme. Note that in the existing literature, \citet{ForsythV2014} considered only the strategy $\Gamma^L$ and computed the ``total value'', equivalent to $V(0,\bX(0);\Gamma^L)+M(0,\bX(0);\Gamma^L)$, and obtained the fair fee rate by requiring this total value to be equal to $W(0)$, the initial investment, which is the same as requiring the initial liability to be equal to 0 as indicated by (\ref{fairalpha}).

\subsection{Results and implications}
\begin{figure}[p!]
	\centering
	\makebox[0pt][c]{
		\begin{tabular}{cc}
			\includegraphics[width=0.5\textwidth]{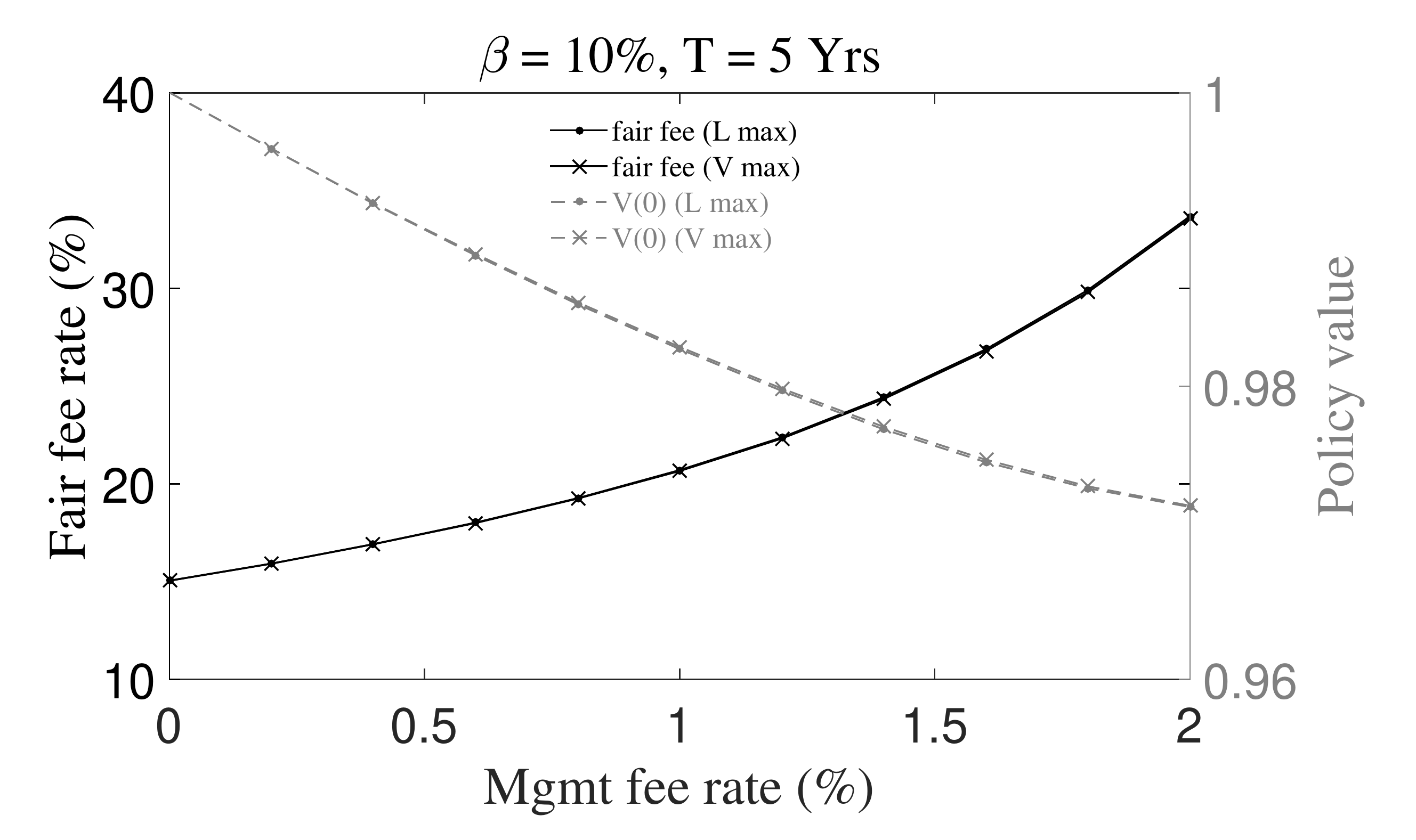}&
			\includegraphics[width=0.5\textwidth]{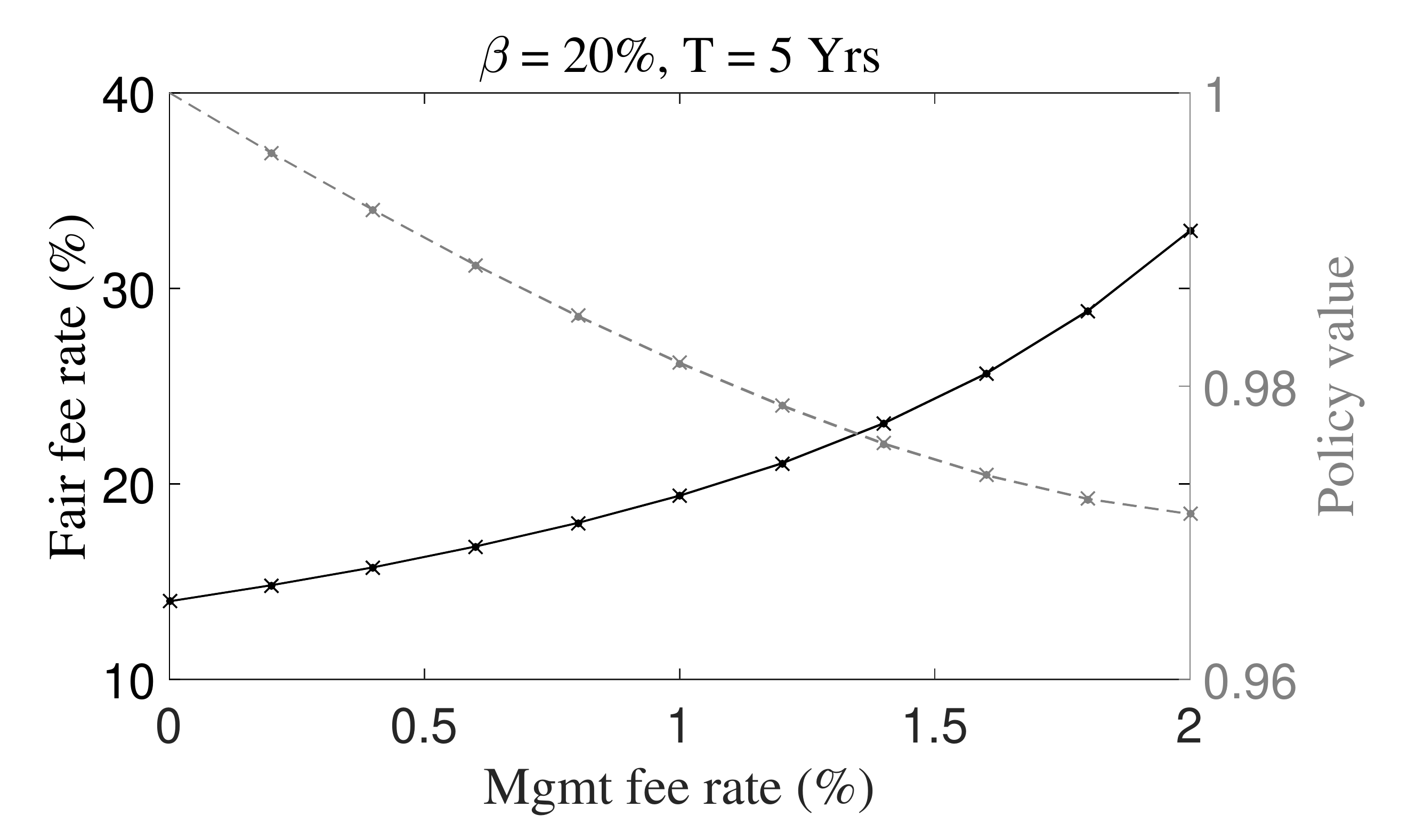}\\
			\includegraphics[width=0.5\textwidth]{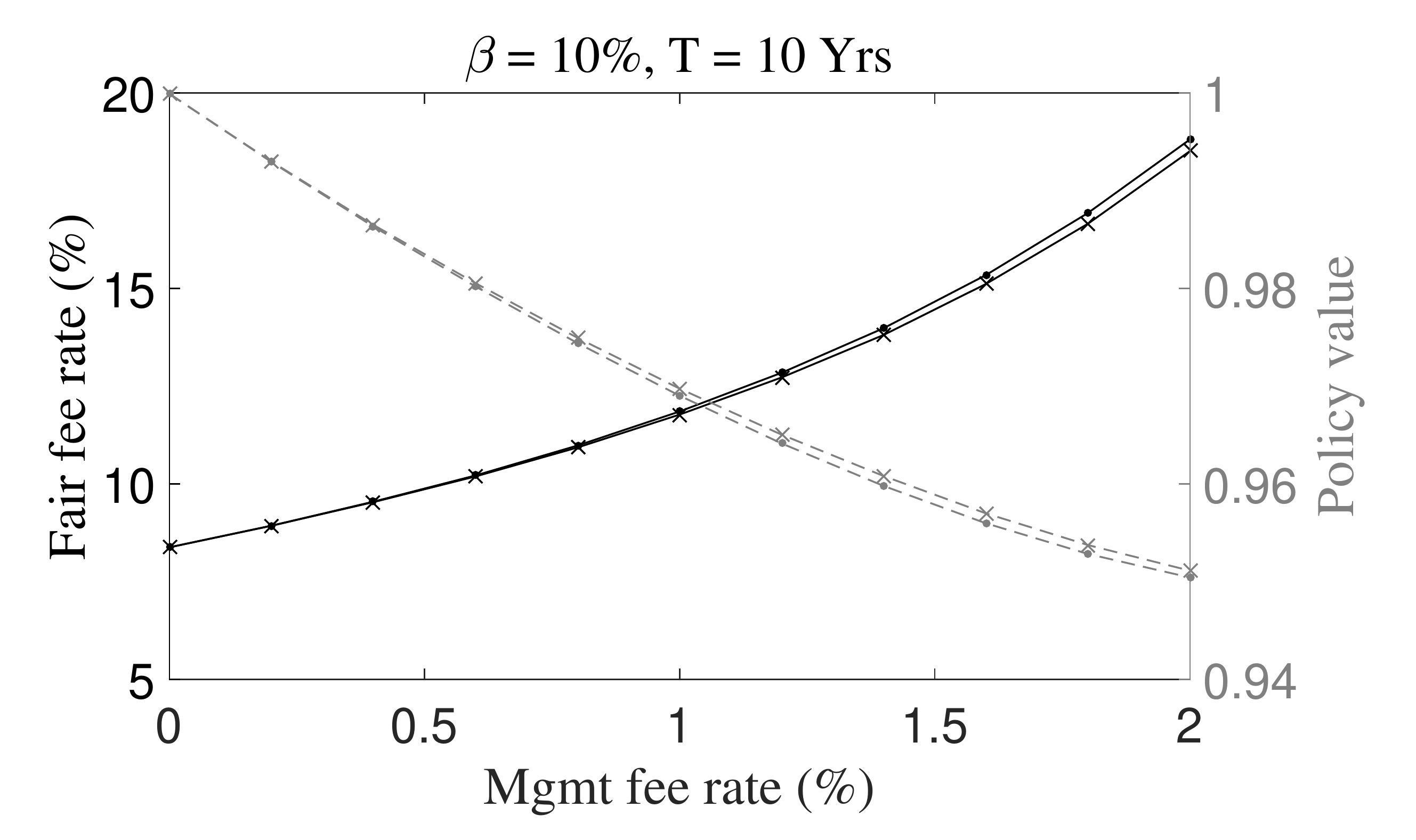}&
			\includegraphics[width=0.5\textwidth]{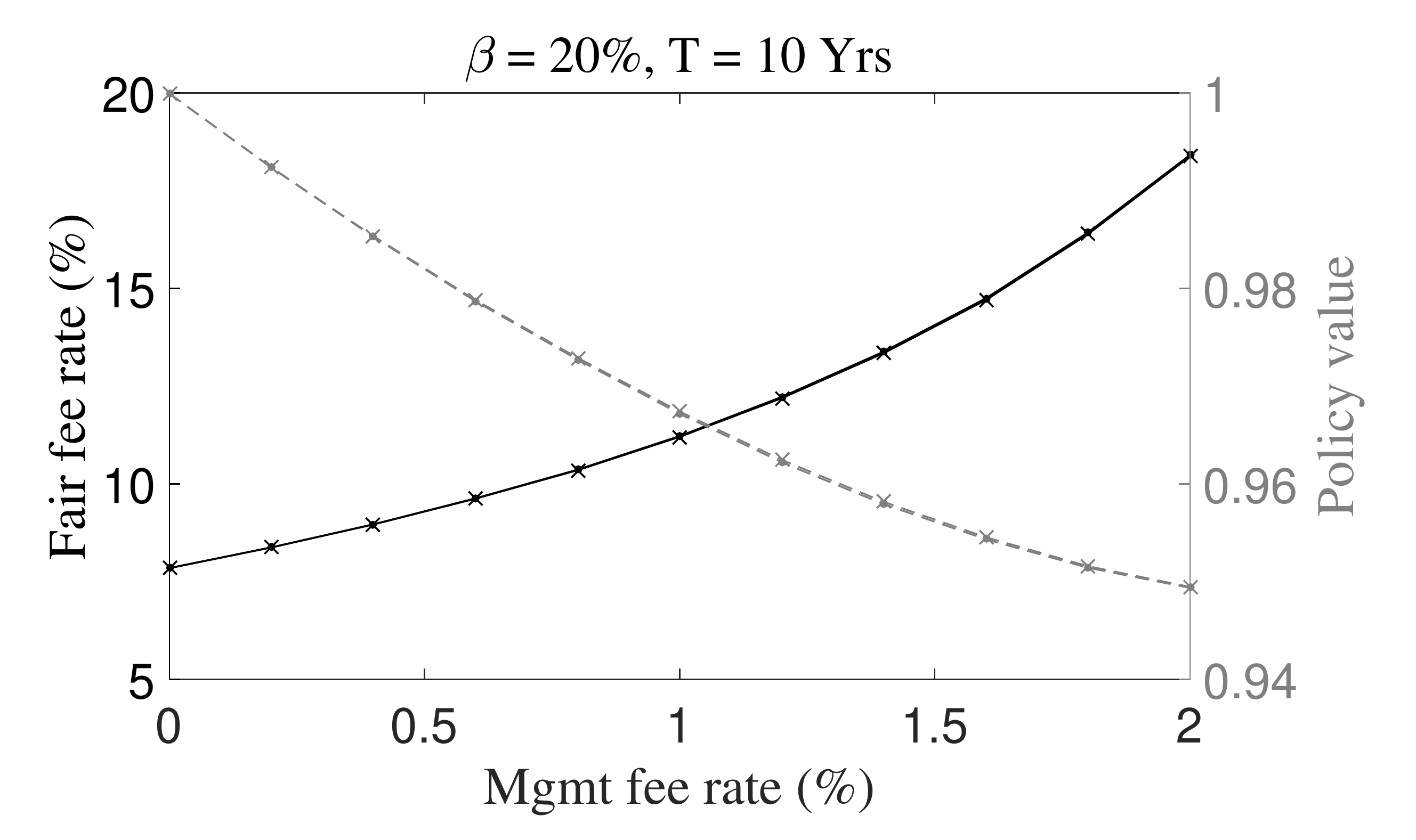}\\
			\includegraphics[width=0.5\textwidth]{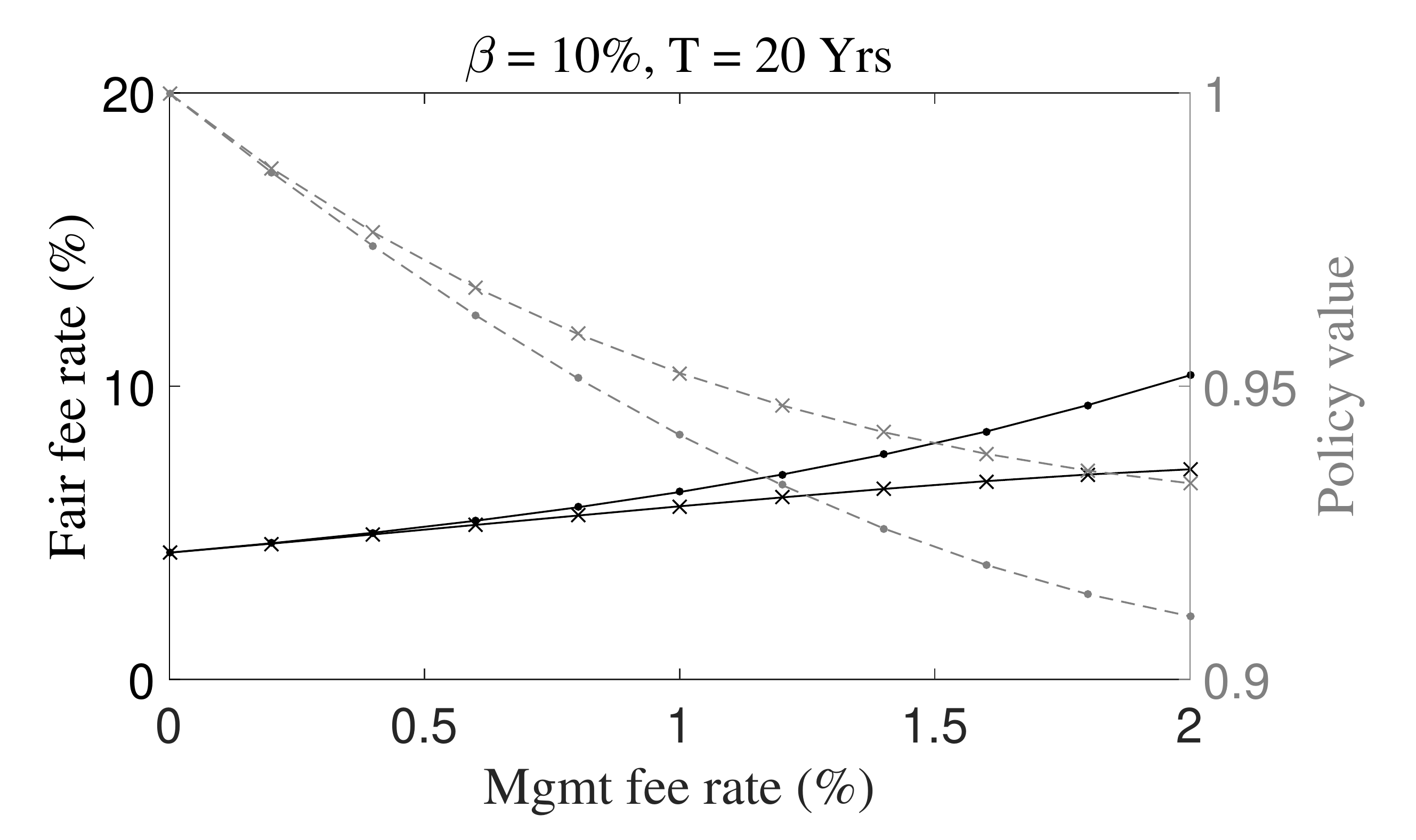}&
			\includegraphics[width=0.5\textwidth]{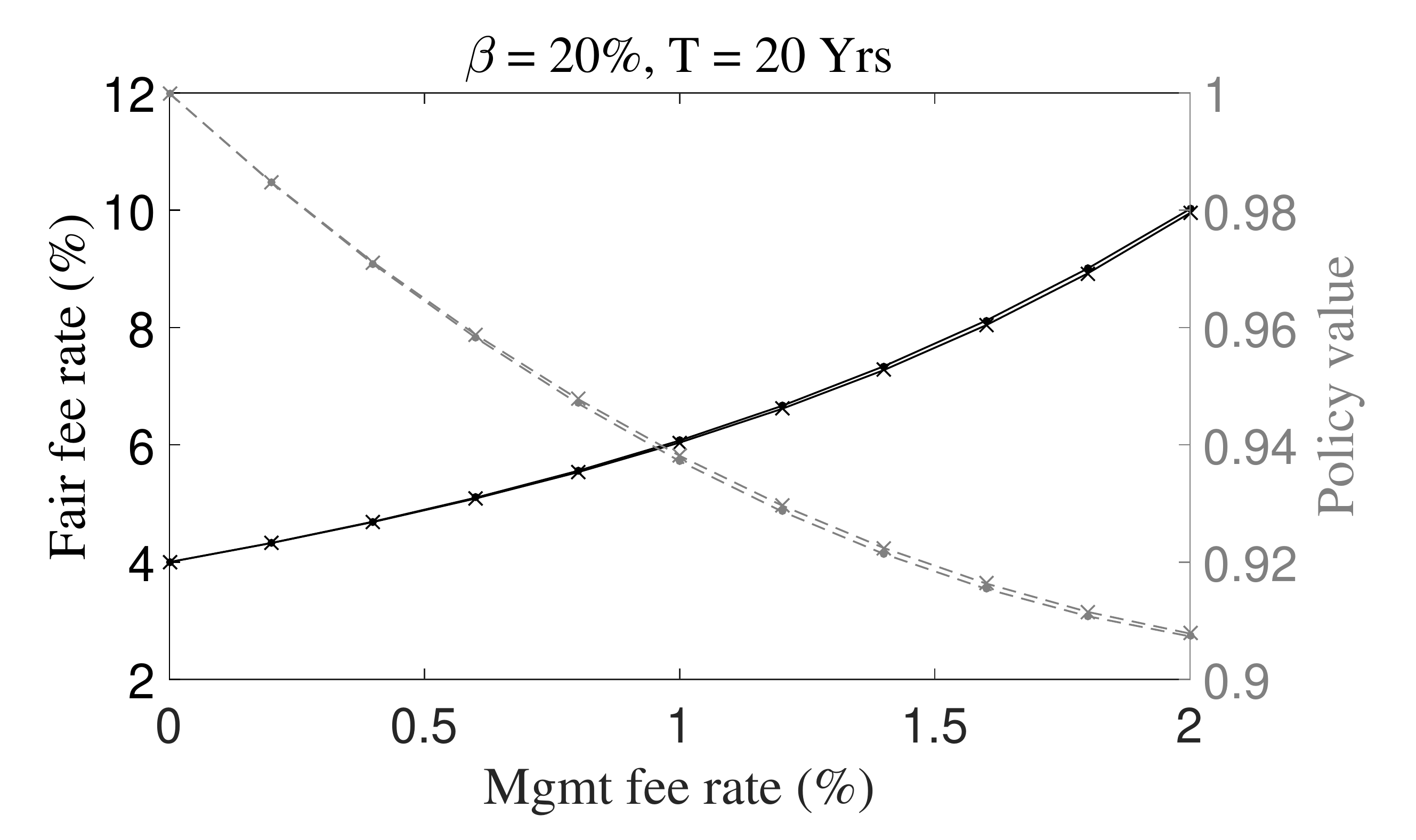}\\
		\end{tabular}
	}
	\caption{Fair insurance fee rates and policy values as a function of management fee rates $\alpha_{\rm m}$ for risk-free rate $r=1\%$ and volatility $\sigma=30\%$, for penalty rates $\beta=10\%\,,20\%$ and maturities $T=5\,,10\,,20$ years. The left axis and dark plots refer to the fair fees in percentage; The right axis and gray plots refer to the policy values. Legends across all plots are shown in the upper left panel.}
	\label{ffee1}
\end{figure}

\begin{figure}[p!]
	\centering
	\makebox[0pt][c]{
		\begin{tabular}{cc}
			\includegraphics[width=0.5\textwidth]{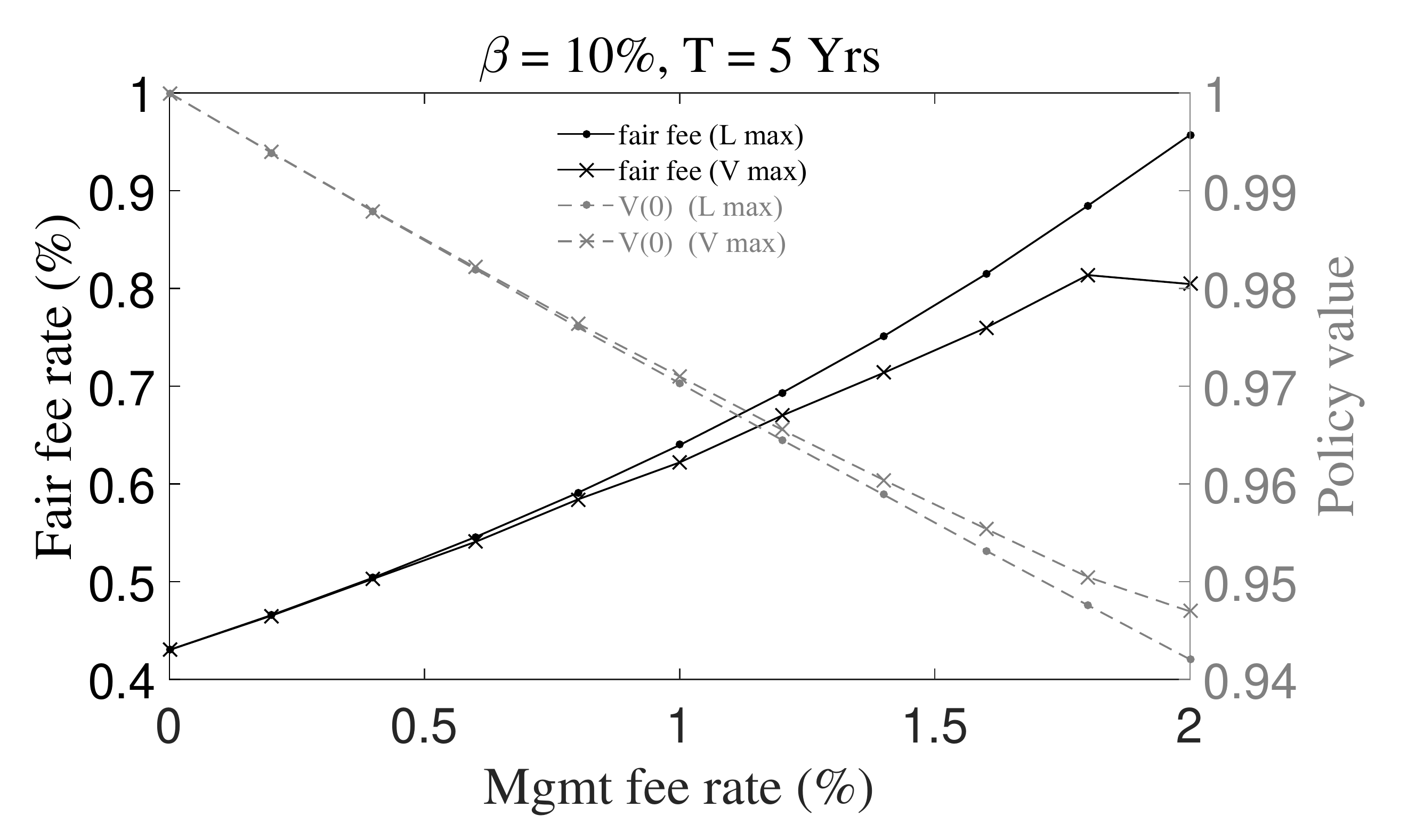}&
			\includegraphics[width=0.5\textwidth]{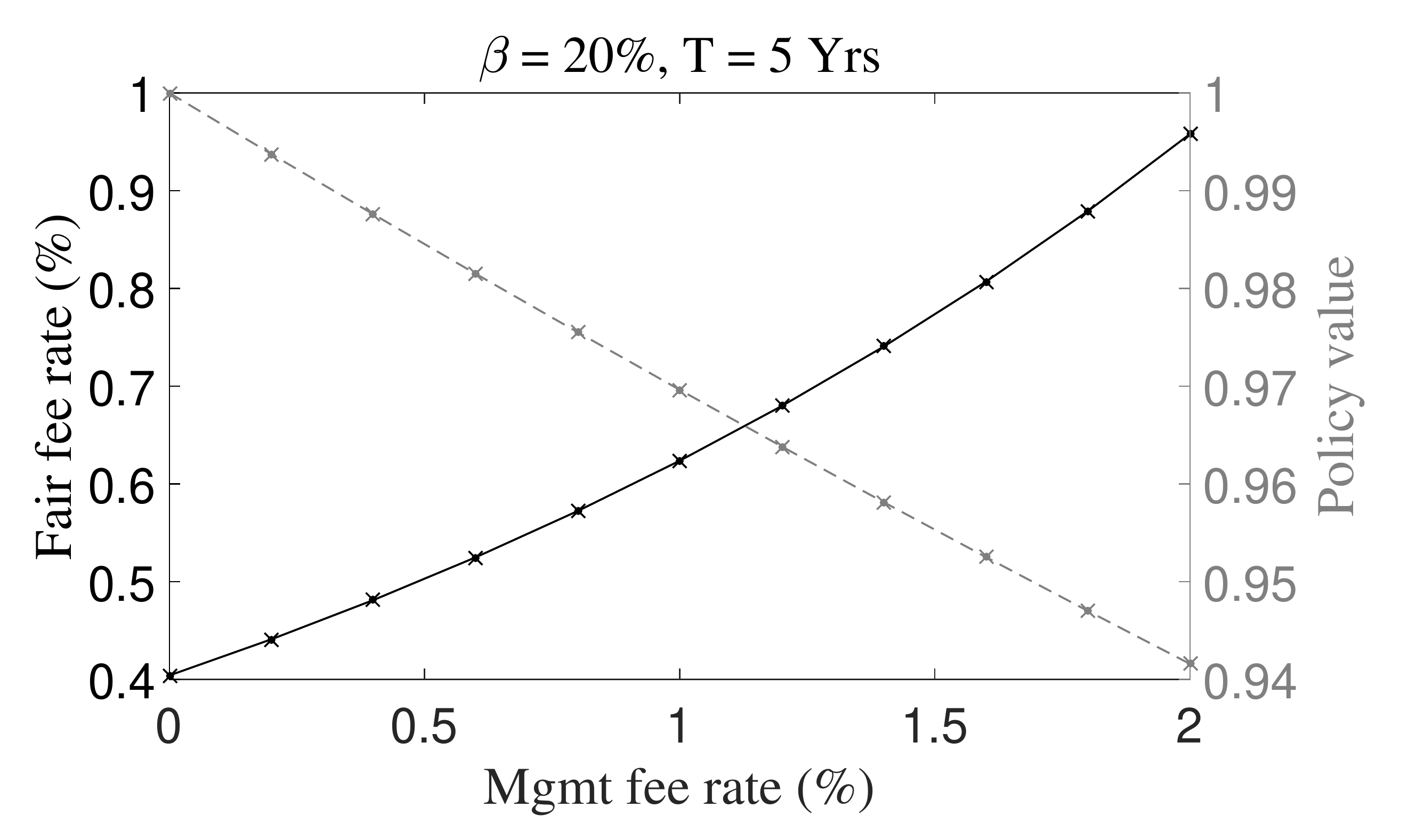}\\
			\includegraphics[width=0.5\textwidth]{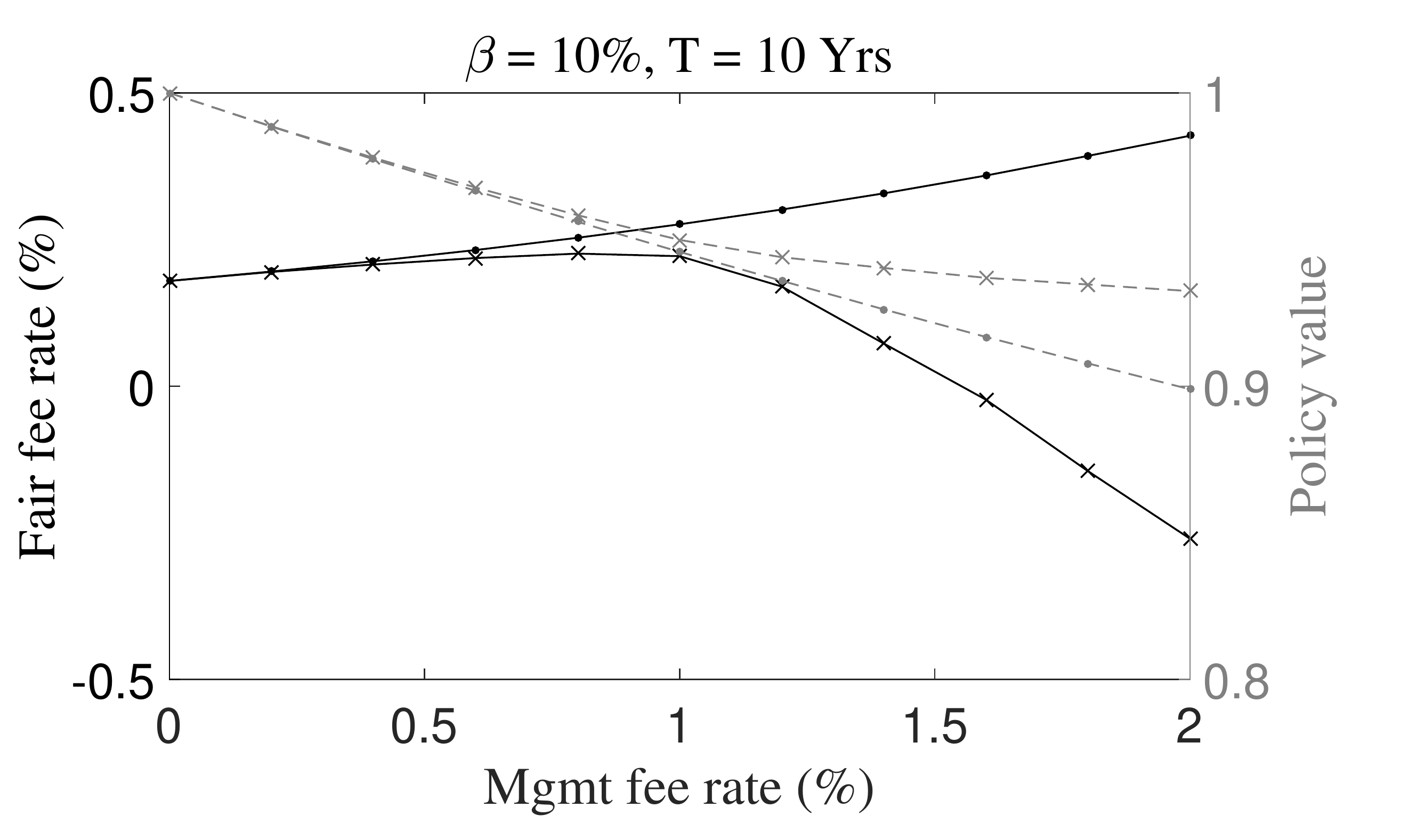}&
			\includegraphics[width=0.5\textwidth]{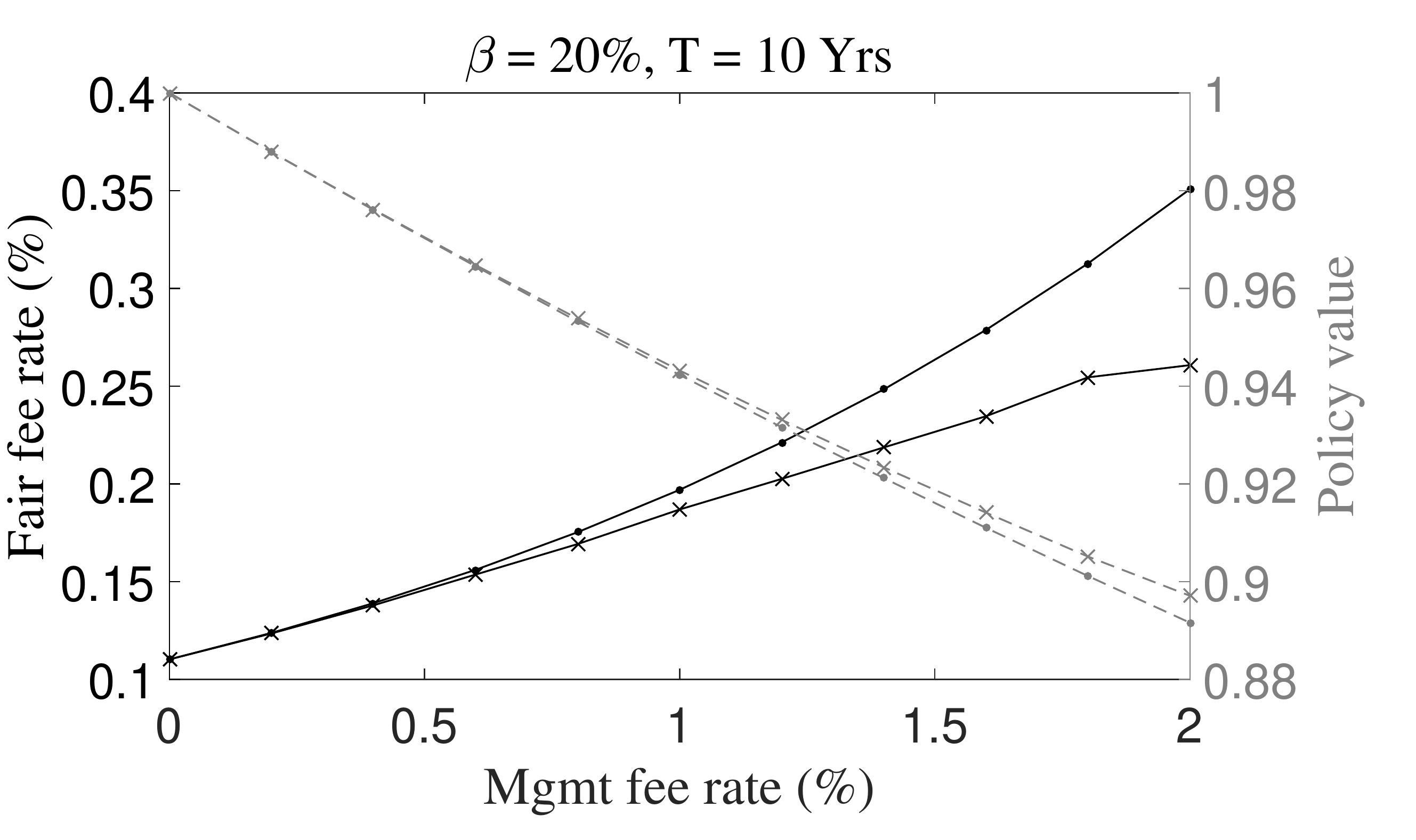}\\
			\includegraphics[width=0.5\textwidth]{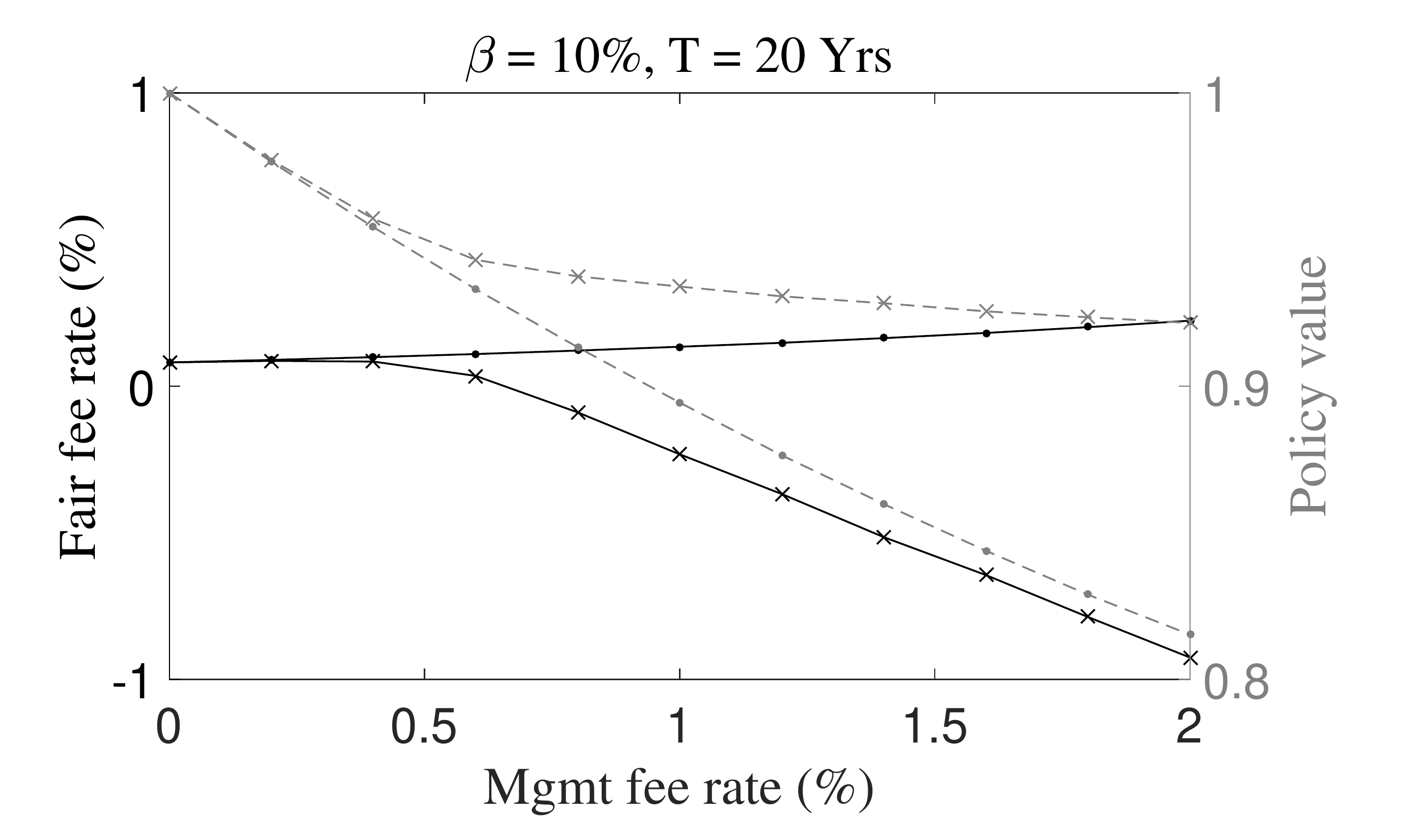}&
			\includegraphics[width=0.5\textwidth]{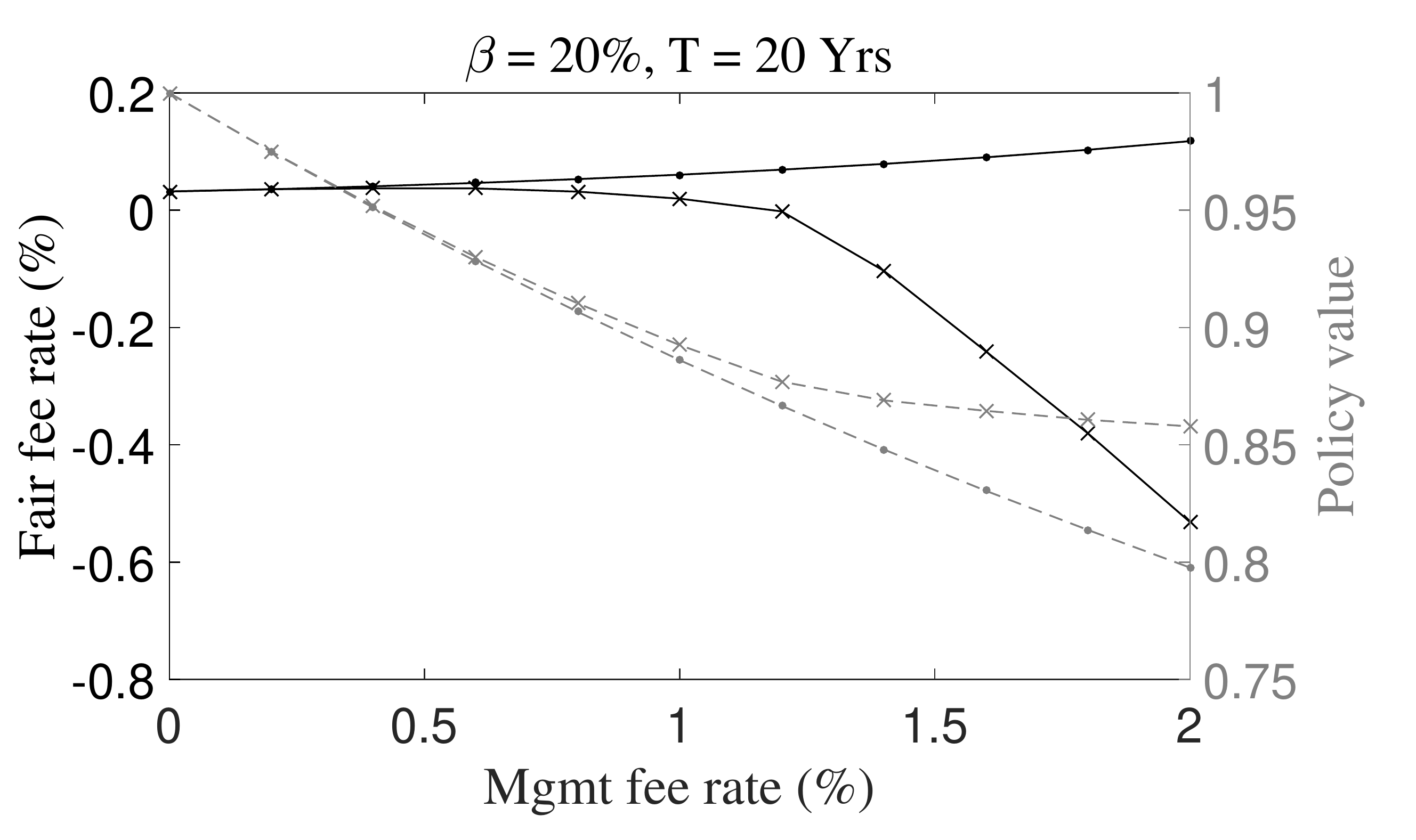}\\
		\end{tabular}
	}
	
	\caption{Fair insurance fee rates and policy values as a function of management fee rates $\alpha_{\rm m}$  for risk-free rate $r=5\%$ and volatility $\sigma=10\%$, for penalty rates $\beta=10\%\,,20\%$ and maturities $T=5\,,10\,,20$ years. The left axis and dark plots refer to the fair fees in percentage; The right axis and gray plots refer to the policy values. Legends across all plots are shown in the upper left panel.}
	\label{ffee2}
\end{figure}

The fair fees and corresponding total policy values are shown in Figures \ref{ffee1} and \ref{ffee2} for two market conditions: a low return market with high volatility ($r=1\%,\sigma=30\%$) and a high return market with low volatility ($r=5\%,\sigma=10\%$), respectively. Fair fee rates obtained for all market conditions and contract parameters can be found in Tables \ref{tffeelm} and \ref{tffeeqm}. The corresponding policy values are listed in Tables \ref{tpvlm} and \ref{tpvqm}.

The first observation to note from these numerical results is that the fair fee rate implied by the liability maximization strategy is always higher, and the corresponding policyholder's total value always lower, than those implied by the value maximization strategy, unless management fees are absent, in which case these quantities are equal. These are to be expected from the definitions of the two strategies.

We also observe from these numerical results that, under the market condition of low return with high volatility, a much higher insurance fee rate is required than under the market condition of high return with low volatility, for the obvious reason that under adverse market conditions, the guarantee is more valuable. Moreover, a higher penalty rate results in a lower insurance fee since a higher penalty rate discourages the policyholder from making more desirable withdrawals that exceed the contracted values.

Furthermore, the results show that under most market conditions or contract specifications, the fair insurance fee rate obtained is highly sensitive to the management fee rate regardless of the withdrawal strategies, as seen from Figures \ref{ffee1} and \ref{ffee2}. In particular, the fair fee rate implied by the liability maximization strategy always increases with the management fee rate, since the management fees cause the wealth account to decrease, leading to higher liability for the insurer to fulfill. On the other hand, the fair fee rate implied by the value maximization strategy first increases then decreases with the management fee rate, since at high management fee rates, a rational policyholder tends to withdraw more and early to avoid the management fees, which in turn reduces the liability and generates more penalty incomes for the insurer. 

A major insight from the numerical results is that with increasing management fees, the value maximization withdrawals of a rational policyholder deviates more from the liability maximization withdrawals assumed by the insurer. In particular, it is seen by examining Figures \ref{ffee1} and \ref{ffee2} that the fair fee rates implied by the two strategies differ more significantly under the following conditions:
\begin{itemize}
	\item longer maturity $T$,
	\item lower penalty rate $\beta$,
	\item higher index return $r$, and
	\item higher management fee rate $\alpha_{\rm m}$.
\end{itemize}
Moreover, careful examination of results listed in Tables \ref{tffeelm} and \ref{tffeeqm} reveals that the index volatility $\sigma$ does not seem to contribute significantly to this discrepancy. These observations are intuitively reasonable: The contributors listed above all imply that the total management fees $M(0)$ will be higher. There are more incentives to withdraw early to achieve more values in the form of reduced management fees. The corresponding differences between the policyholder's values follow similar patterns. Of particular interest is that in some cases, as shown in Figure \ref{ffee2}, the fair fee rate implied by maximizing policyholder's value can become negative. This implies that the policyholder would want to withdraw more and early due to high management fees to such an extent, that the penalties incurred exceed the total value of the GMWB rider. On the other hand, the fair fee rate implied by maximizing the liability is always positive.

\section{Conclusions}
\label{conc}
Determining accurate hedging costs of VA guarantees is a significant issue for VA providers. While the presence of management fees is typically ignored in the VA literature, it was demonstrated in this article that the impact of management fees on the pricing of GMWB contract is significant. As a form of market friction similar to tax consideration, management fees can affect policyholders' withdrawal behaviors, causing large deviations from the ``optimal" (liability maximization) withdrawal behaviors often assumed in the literature.

Two different policyholder's withdrawal strategies were considered: liability maximization and value maximization when management fees are present. We demonstrated that these two  withdrawal strategies imply different fair insurance fee rates, where  maximizing policy value implies lower fair fees than those implied by maximizing liability, or equivalently, maximizing the ``total value'' of the contract, which represents the maximal hedging costs from the insurer's perspective.

More importantly, we quantitatively demonstrated that the difference between the initial investment and the total value of the policyholder is precisely the total value of the management fees, which is also the cause of the discrepancy between the two withdrawal strategies considered in this article. The two strategies coincide when management fees are absent.  We identified a number of factors that contribute to this discrepancy through a series of illustrating numerical experiments. 
Our findings identify the management fees as a potential cause of discrepancy between the fair fee rates implied by the liability maximization strategy, often assumed from the insurer's perspective for VA pricing, and the prevailing market rates for VA contracts with GMWB or similar riders.

\section*{Acknowledgement}
This research was supported by the CSIRO-Monash Superannuation Research Cluster, a collaboration among CSIRO, Monash University, Griffith University, the University of Western Australia, the University of Warwick, and stakeholders of the retirement system in the interest of better outcomes for all. This research was also partially supported under the Australian Research Council's Discovery Projects funding scheme (project number: DP160103489). We would like to thank Eckhard Platen and Xiaolin Luo for useful discussions and comments.

\begin{table}[p]
	\centering
	\caption{Fair fee rate $\alpha_{\rm ins}$ ($\%$) based on the liability maximization strategy $\Gamma^L$.}
	\begin{tabular}{rrrr|rrrrrrrrrrr}
		\toprule
		\multicolumn{4}{c|}{Parameters} & \multicolumn{11}{c}{$\alpha_{\rm m}$} \\
		$r(\%)$  & $\sigma(\%)$  & $\beta(\%)$  & $T$  & $0\%$  & $0.2\%$  & $0.4\%$  & $0.6\%$  & $0.8\%$  & $1\%$  & $1.2\%$  & $1.4\%$  & $1.6\%$  & $1.8\%$  & $2\%$ \\ \hline
		\midrule
		1  & 10  & 10  & 5  & 3.08 & 3.48 & 3.96 & 4.59 & 5.41 & 6.65 & 8.66 & 12.18 & 17.67 & 23.53 & 29.92 \\
		&  &  & 10  & 1.66 & 1.92 & 2.25 & 2.66 & 3.21 & 3.97 & 5.08 & 6.77 & 9.26 & 12.17 & 15.31 \\
		&  &  & 20  & 0.82 & 0.97 & 1.17 & 1.43 & 1.77 & 2.22 & 2.83 & 3.68 & 4.81 & 6.19 & 7.71 \\
		&  & 20  & 5  & 3.08 & 3.47 & 3.95 & 4.55 & 5.34 & 6.48 & 8.39 & 12.01 & 17.66 & 23.55 & 29.92 \\
		&  &  & 10  & 1.66 & 1.91 & 2.23 & 2.62 & 3.13 & 3.83 & 4.88 & 6.60 & 9.20 & 12.17 & 15.31 \\
		&  &  & 20  & 0.81 & 0.96 & 1.16 & 1.40 & 1.71 & 2.13 & 2.72 & 3.57 & 4.74 & 6.17 & 7.71 \\
		& 30  & 10  & 5  & 15.05 & 15.92 & 16.92 & 18.02 & 19.27 & 20.70 & 22.37 & 24.43 & 26.88 & 29.91 & 33.69 \\
		&  &  & 10  & 8.38 & 8.93 & 9.55 & 10.22 & 10.98 & 11.85 & 12.84 & 13.98 & 15.34 & 16.93 & 18.81 \\
		&  &  & 20  & 4.32 & 4.64 & 5.00 & 5.41 & 5.87 & 6.39 & 6.98 & 7.66 & 8.45 & 9.35 & 10.37 \\
		&  & 20  & 5  & 13.99 & 14.82 & 15.73 & 16.80 & 18.01 & 19.40 & 21.06 & 23.11 & 25.64 & 28.86 & 32.95 \\
		&  &  & 10  & 7.85 & 8.38 & 8.97 & 9.63 & 10.36 & 11.22 & 12.22 & 13.38 & 14.74 & 16.43 & 18.41 \\
		&  &  & 20  & 4.00 & 4.33 & 4.69 & 5.10 & 5.55 & 6.07 & 6.66 & 7.34 & 8.11 & 9.01 & 10.02 \\
		5  & 10  & 10  & 5  & 0.43 & 0.47 & 0.50 & 0.55 & 0.59 & 0.64 & 0.69 & 0.75 & 0.81 & 0.88 & 0.96 \\
		&  &  & 10  & 0.18 & 0.20 & 0.21 & 0.23 & 0.25 & 0.28 & 0.30 & 0.33 & 0.36 & 0.39 & 0.43 \\
		&  &  & 20  & 0.08 & 0.09 & 0.10 & 0.11 & 0.12 & 0.13 & 0.15 & 0.16 & 0.18 & 0.20 & 0.22 \\
		&  & 20  & 5  & 0.40 & 0.44 & 0.48 & 0.52 & 0.57 & 0.62 & 0.68 & 0.74 & 0.81 & 0.88 & 0.96 \\
		&  &  & 10  & 0.11 & 0.12 & 0.14 & 0.16 & 0.18 & 0.20 & 0.22 & 0.25 & 0.28 & 0.31 & 0.35 \\
		&  &  & 20  & 0.03 & 0.04 & 0.04 & 0.05 & 0.05 & 0.06 & 0.07 & 0.08 & 0.09 & 0.10 & 0.12 \\
		& 30  & 10  & 5  & 5.33 & 5.48 & 5.65 & 5.81 & 5.99 & 6.17 & 6.35 & 6.55 & 6.75 & 6.96 & 7.17 \\
		&  &  & 10  & 2.91 & 3.02 & 3.12 & 3.23 & 3.35 & 3.47 & 3.60 & 3.73 & 3.86 & 4.01 & 4.16 \\
		&  &  & 20  & 1.58 & 1.65 & 1.74 & 1.82 & 1.91 & 2.00 & 2.10 & 2.21 & 2.32 & 2.43 & 2.55 \\
		&  & 20  & 5  & 4.97 & 5.13 & 5.28 & 5.44 & 5.61 & 5.79 & 5.97 & 6.15 & 6.35 & 6.55 & 6.76 \\
		&  &  & 10  & 2.27 & 2.35 & 2.43 & 2.52 & 2.61 & 2.71 & 2.81 & 2.91 & 3.02 & 3.13 & 3.25 \\
		&  &  & 20  & 1.08 & 1.13 & 1.19 & 1.24 & 1.31 & 1.37 & 1.43 & 1.50 & 1.58 & 1.65 & 1.73 \\
		\bottomrule
	\end{tabular}
	\label{tffeelm}
\end{table}

\begin{table}[p]
	\centering
	\caption{Fair fee rate $\alpha_{\rm ins}$ ($\%$) based on the policy value maximization strategy $\Gamma^V$.}
	\begin{tabular}{rrrr|rrrrrrrrrrr}
		\toprule
		\multicolumn{4}{c|}{Parameters} & \multicolumn{11}{c}{$\alpha_{\rm m}$} \\
		$r(\%)$  & $\sigma(\%)$  & $\beta(\%)$  & $T$  & $0\%$  & $0.2\%$  & $0.4\%$  & $0.6\%$  & $0.8\%$  & $1\%$  & $1.2\%$  & $1.4\%$  & $1.6\%$  & $1.8\%$  & $2\%$ \\ \hline
		\midrule
		1 & 10  & 10  & 5 & 3.08 & 3.47 & 3.96 & 4.57 & 5.36 & 6.57 & 8.55 & 12.12 & 17.66 & 23.55 & 29.93 \\
		&  &  & 10  & 1.66 & 1.92 & 2.23 & 2.61 & 3.13 & 3.81 & 4.86 & 6.44 & 8.99 & 12.11 & 15.30 \\
		&  &  & 20  & 0.82 & 0.96 & 1.09 & 1.21 & 1.33 & 1.44 & 1.50 & 1.59 & 1.65 & 1.70 & 1.81 \\
		&  & 20  & 5 & 3.08 & 3.47 & 3.95 & 4.55 & 5.34 & 6.48 & 8.38 & 12.01 & 17.66 & 23.55 & 29.92 \\
		&  &  & 10  & 1.66 & 1.91 & 2.23 & 2.62 & 3.12 & 3.80 & 4.84 & 6.56 & 9.19 & 12.17 & 15.31 \\
		&  &  & 20  & 0.81 & 0.96 & 1.16 & 1.39 & 1.67 & 2.05 & 2.60 & 3.43 & 4.65 & 6.14 & 7.70 \\
		& 30  & 10  & 5  & 15.05 & 15.92 & 16.91 & 18.00 & 19.25 & 20.66 & 22.32 & 24.34 & 26.79 & 29.80 & 33.58 \\
		&  &  & 10  & 8.38 & 8.93 & 9.53 & 10.19 & 10.93 & 11.77 & 12.72 & 13.81 & 15.11 & 16.65 & 18.51 \\
		&  &  & 20  & 4.32 & 4.63 & 4.94 & 5.27 & 5.59 & 5.90 & 6.20 & 6.49 & 6.76 & 6.99 & 7.16 \\
		&  & 20  & 5  & 13.99 & 14.82 & 15.73 & 16.80 & 18.00 & 19.39 & 21.04 & 23.09 & 25.62 & 28.84 & 32.93 \\
		&  &  & 10  & 7.85 & 8.38 & 8.96 & 9.62 & 10.35 & 11.20 & 12.19 & 13.34 & 14.70 & 16.38 & 18.37 \\
		&  &  & 20  & 4.00 & 4.32 & 4.68 & 5.08 & 5.53 & 6.04 & 6.61 & 7.27 & 8.04 & 8.92 & 9.95 \\
		5  & 10  & 10  & 5  & 0.43 & 0.47 & 0.50 & 0.54 & 0.58 & 0.62 & 0.67 & 0.71 & 0.76 & 0.81 & 0.80 \\
		&  &  & 10  & 0.18 & 0.19 & 0.21 & 0.22 & 0.23 & 0.22 & 0.17 & 0.07 & -0.02 & -0.14 & -0.26 \\
		&  &  & 20  & 0.08 & 0.09 & 0.08 & 0.04 & -0.09 & -0.23 & -0.37 & -0.51 & -0.64 & -0.79 & -0.93 \\
		&  & 20  & 5  & 0.40 & 0.44 & 0.48 & 0.52 & 0.57 & 0.62 & 0.68 & 0.74 & 0.81 & 0.88 & 0.96 \\
		&  &  & 10  & 0.11 & 0.12 & 0.14 & 0.15 & 0.17 & 0.19 & 0.20 & 0.22 & 0.23 & 0.25 & 0.26 \\
		&  &  & 20  & 0.03 & 0.04 & 0.04 & 0.04 & 0.03 & 0.02 & -0.00 & -0.10 & -0.24 & -0.38 & -0.53 \\
		& 30  & 10  & 5  & 5.33 & 5.48 & 5.64 & 5.79 & 5.95 & 6.11 & 6.27 & 6.43 & 6.60 & 6.75 & 6.94 \\
		&  &  & 10  & 2.91 & 3.01 & 3.10 & 3.18 & 3.26 & 3.34 & 3.41 & 3.47 & 3.53 & 3.58 & 3.60 \\
		&  &  & 20  & 1.58 & 1.64 & 1.68 & 1.72 & 1.74 & 1.75 & 1.75 & 1.74 & 1.74 & 1.73 & 1.71 \\
		&  & 20  & 5  & 4.97 & 5.13 & 5.28 & 5.44 & 5.61 & 5.78 & 5.96 & 6.14 & 6.32 & 6.51 & 6.70 \\
		&  &  & 10  & 2.27 & 2.35 & 2.43 & 2.51 & 2.59 & 2.67 & 2.74 & 2.82 & 2.88 & 2.94 & 3.00 \\
		&  &  & 20  & 1.08 & 1.13 & 1.18 & 1.22 & 1.23 & 1.24 & 1.23 & 1.21 & 1.18 & 1.14 & 1.09 \\
		\bottomrule
	\end{tabular}
	\label{tffeeqm}
\end{table}

\begin{table}[p]
	\centering
	\caption{Total policy value $V(0,\bX(0);\Gamma^L)$ based on the liability maximization strategy $\Gamma^L$.}
	\begin{tabular}{rrrr|rrrrrrrrrrr}
		\toprule
		\multicolumn{4}{c|}{Parameters} & \multicolumn{11}{c}{$\alpha_{\rm m}$} \\
		$r(\%)$  & $\sigma(\%)$  & $\beta(\%)$ & $T$  & $0\%$  & $0.2\%$  & $0.4\%$  & $0.6\%$  & $0.8\%$  & $1\%$  & $1.2\%$  & $1.4\%$  & $1.6\%$  & $1.8\%$  & $2\%$ \\ \hline
		\midrule
		1  & 10  & 10  & 5  & 1.00 & 0.99 & 0.99 & 0.98 & 0.98 & 0.98 & 0.97 & 0.97 & 0.97 & 0.97 & 0.97 \\
		&  &  & 10  & 1.00 & 0.99 & 0.98 & 0.97 & 0.97 & 0.96 & 0.95 & 0.95 & 0.95 & 0.95 & 0.95 \\
		&  &  & 20  & 1.00 & 0.98 & 0.96 & 0.95 & 0.94 & 0.92 & 0.92 & 0.91 & 0.90 & 0.90 & 0.90 \\
		&  & 20  & 5  & 1.00 & 0.99 & 0.99 & 0.98 & 0.98 & 0.98 & 0.97 & 0.97 & 0.97 & 0.97 & 0.97 \\
		&  &  & 10  & 1.00 & 0.99 & 0.98 & 0.97 & 0.96 & 0.96 & 0.95 & 0.95 & 0.95 & 0.95 & 0.95 \\
		&  &  & 20  & 1.00 & 0.98 & 0.96 & 0.95 & 0.93 & 0.92 & 0.91 & 0.91 & 0.90 & 0.90 & 0.90 \\
		& 30  & 10  & 5  & 1.00 & 1.00 & 0.99 & 0.99 & 0.99 & 0.98 & 0.98 & 0.98 & 0.97 & 0.97 & 0.97 \\
		&  &  & 10  & 1.00 & 0.99 & 0.99 & 0.98 & 0.97 & 0.97 & 0.96 & 0.96 & 0.96 & 0.95 & 0.95 \\
		&  &  & 20  & 1.00 & 0.99 & 0.97 & 0.96 & 0.95 & 0.94 & 0.93 & 0.93 & 0.92 & 0.91 & 0.91 \\
		&  & 20  & 5  & 1.00 & 1.00 & 0.99 & 0.99 & 0.98 & 0.98 & 0.98 & 0.98 & 0.97 & 0.97 & 0.97 \\
		&  &  & 10  & 1.00 & 0.99 & 0.99 & 0.98 & 0.97 & 0.97 & 0.96 & 0.96 & 0.95 & 0.95 & 0.95 \\
		&  &  & 20  & 1.00 & 0.98 & 0.97 & 0.96 & 0.95 & 0.94 & 0.93 & 0.92 & 0.92 & 0.91 & 0.91 \\
		5  & 10  & 10  & 5  & 1.00 & 0.99 & 0.99 & 0.98 & 0.98 & 0.97 & 0.96 & 0.96 & 0.95 & 0.95 & 0.94 \\
		&  &  & 10  & 1.00 & 0.99 & 0.98 & 0.97 & 0.96 & 0.95 & 0.94 & 0.93 & 0.92 & 0.91 & 0.90 \\
		&  &  & 20  & 1.00 & 0.98 & 0.95 & 0.93 & 0.91 & 0.89 & 0.88 & 0.86 & 0.84 & 0.83 & 0.82 \\
		&  & 20  & 5  & 1.00 & 0.99 & 0.99 & 0.98 & 0.98 & 0.97 & 0.96 & 0.96 & 0.95 & 0.95 & 0.94 \\
		&  &  & 10  & 1.00 & 0.99 & 0.98 & 0.96 & 0.95 & 0.94 & 0.93 & 0.92 & 0.91 & 0.90 & 0.89 \\
		&  &  & 20  & 1.00 & 0.97 & 0.95 & 0.93 & 0.91 & 0.89 & 0.87 & 0.85 & 0.83 & 0.81 & 0.80 \\
		& 30  & 10  & 5  & 1.00 & 1.00 & 0.99 & 0.99 & 0.98 & 0.98 & 0.97 & 0.97 & 0.96 & 0.96 & 0.96 \\
		&  &  & 10  & 1.00 & 0.99 & 0.98 & 0.98 & 0.97 & 0.96 & 0.95 & 0.95 & 0.94 & 0.93 & 0.93 \\
		&  &  & 20  & 1.00 & 0.98 & 0.97 & 0.96 & 0.94 & 0.93 & 0.92 & 0.91 & 0.90 & 0.89 & 0.88 \\
		&  & 20  & 5  & 1.00 & 0.99 & 0.99 & 0.98 & 0.98 & 0.97 & 0.97 & 0.96 & 0.96 & 0.95 & 0.95 \\
		&  &  & 10  & 1.00 & 0.99 & 0.98 & 0.97 & 0.96 & 0.95 & 0.94 & 0.93 & 0.92 & 0.91 & 0.91 \\
		&  &  & 20  & 1.00 & 0.98 & 0.96 & 0.94 & 0.92 & 0.90 & 0.89 & 0.87 & 0.86 & 0.84 & 0.83 \\
		\bottomrule
	\end{tabular}
	\label{tpvlm}
\end{table}

\begin{table}[p]
	\centering
	\caption{Total policy value $V(0,\bX(0);\Gamma^V)$ based on the policy value maximization strategy $\Gamma^V$.}
	\begin{tabular}{rrrr|rrrrrrrrrrr}
		\toprule
		\multicolumn{4}{c|}{Parameters} & \multicolumn{11}{c}{$\alpha_{\rm m}$} \\
		$r(\%)$  & $\sigma(\%)$  & $\beta(\%)$  & $T$  & $0\%$  & $0.2\%$  & $0.4\%$  & $0.6\%$  & $0.8\%$  & $1\%$  & $1.2\%$  & $1.4\%$  & $1.6\%$  & $1.8\%$  & $2\%$ \\ \hline
		\midrule
		1  & 10  & 10  & 5  & 1.00 & 0.99 & 0.99 & 0.98 & 0.98 & 0.98 & 0.97 & 0.97 & 0.97 & 0.97 & 0.97 \\
		&  &  & 10  & 1.00 & 0.99 & 0.98 & 0.97 & 0.97 & 0.96 & 0.95 & 0.95 & 0.95 & 0.95 & 0.95 \\
		&  &  & 20  & 1.00 & 0.98 & 0.97 & 0.96 & 0.95 & 0.95 & 0.94 & 0.94 & 0.94 & 0.93 & 0.93 \\
		&  & 20  & 5  & 1.00 & 0.99 & 0.99 & 0.98 & 0.98 & 0.98 & 0.97 & 0.97 & 0.97 & 0.97 & 0.97 \\
		&  &  & 10  & 1.00 & 0.99 & 0.98 & 0.97 & 0.96 & 0.96 & 0.95 & 0.95 & 0.95 & 0.95 & 0.95 \\
		&  &  & 20  & 1.00 & 0.98 & 0.96 & 0.95 & 0.93 & 0.92 & 0.91 & 0.91 & 0.90 & 0.90 & 0.90 \\
		& 30  & 10  & 5  & 1.00 & 1.00 & 0.99 & 0.99 & 0.99 & 0.98 & 0.98 & 0.98 & 0.97 & 0.97 & 0.97 \\
		&  &  & 10  & 1.00 & 0.99 & 0.99 & 0.98 & 0.97 & 0.97 & 0.97 & 0.96 & 0.96 & 0.95 & 0.95 \\
		&  &  & 20  & 1.00 & 0.99 & 0.98 & 0.97 & 0.96 & 0.95 & 0.95 & 0.94 & 0.94 & 0.94 & 0.93 \\
		&  & 20  & 5  & 1.00 & 1.00 & 0.99 & 0.99 & 0.98 & 0.98 & 0.98 & 0.98 & 0.97 & 0.97 & 0.97 \\
		&  &  & 10  & 1.00 & 0.99 & 0.99 & 0.98 & 0.97 & 0.97 & 0.96 & 0.96 & 0.95 & 0.95 & 0.95 \\
		&  &  & 20  & 1.00 & 0.98 & 0.97 & 0.96 & 0.95 & 0.94 & 0.93 & 0.92 & 0.92 & 0.91 & 0.91 \\
		5  & 10  & 10  & 5  & 1.00 & 0.99 & 0.99 & 0.98 & 0.98 & 0.97 & 0.97 & 0.96 & 0.96 & 0.95 & 0.95 \\
		&  &  & 10  & 1.00 & 0.99 & 0.98 & 0.97 & 0.96 & 0.95 & 0.94 & 0.94 & 0.94 & 0.93 & 0.93 \\
		&  &  & 20  & 1.00 & 0.98 & 0.96 & 0.94 & 0.94 & 0.93 & 0.93 & 0.93 & 0.93 & 0.92 & 0.92 \\
		&  & 20  & 5  & 1.00 & 0.99 & 0.99 & 0.98 & 0.98 & 0.97 & 0.96 & 0.96 & 0.95 & 0.95 & 0.94 \\
		&  &  & 10  & 1.00 & 0.99 & 0.98 & 0.96 & 0.95 & 0.94 & 0.93 & 0.92 & 0.91 & 0.91 & 0.90 \\
		&  &  & 20  & 1.00 & 0.97 & 0.95 & 0.93 & 0.91 & 0.89 & 0.88 & 0.87 & 0.86 & 0.86 & 0.86 \\
		& 30  & 10  & 5  & 1.00 & 1.00 & 0.99 & 0.99 & 0.98 & 0.98 & 0.97 & 0.97 & 0.97 & 0.96 & 0.96 \\
		&  &  & 10  & 1.00 & 0.99 & 0.98 & 0.98 & 0.97 & 0.97 & 0.96 & 0.96 & 0.95 & 0.95 & 0.95 \\
		&  &  & 20  & 1.00 & 0.99 & 0.97 & 0.97 & 0.96 & 0.95 & 0.94 & 0.94 & 0.94 & 0.93 & 0.93 \\
		&  & 20  & 5  & 1.00 & 0.99 & 0.99 & 0.98 & 0.98 & 0.97 & 0.97 & 0.96 & 0.96 & 0.95 & 0.95 \\
		&  &  & 10  & 1.00 & 0.99 & 0.98 & 0.97 & 0.96 & 0.95 & 0.94 & 0.94 & 0.93 & 0.92 & 0.92 \\
		&  &  & 20  & 1.00 & 0.98 & 0.96 & 0.94 & 0.93 & 0.92 & 0.91 & 0.90 & 0.89 & 0.88 & 0.88 \\
		\bottomrule
	\end{tabular}
	\label{tpvqm}
\end{table}

\clearpage
\section*{References}

\bibliographystyle{elsart-harv}
\bibliography{gmwxb}

\begin{thebibliography}{26}
\expandafter\ifx\csname natexlab\endcsname\relax\def\natexlab#1{#1}\fi
\expandafter\ifx\csname url\endcsname\relax
  \def\url#1{\texttt{#1}}\fi
\expandafter\ifx\csname urlprefix\endcsname\relax\def\urlprefix{URL }\fi

\bibitem[{Bacinello and Ortu(1996)}]{BacinelloOr96}
Bacinello, A.~R., Ortu, F., 1996. Fixed income linked life insurance policies
  with minimum guarantees: {P}ricing models and numerical results. European
  Journal of Operational Research 91~(2), 235--249.

\bibitem[{Bauer et~al.(2008)Bauer, Kling, and Russ}]{bauerKR2008}
Bauer, D., Kling, A., Russ, J., 2008. A universal pricing framework for
  guaranteed minimum benefits in variable annuities. ASTIN Bulletin 38~(2),
  621--651.

\bibitem[{B\'{e}langer et~al.(2009)B\'{e}langer, Forsyth, and
  Labahn}]{BelangerFoLa09}
B\'{e}langer, A.~C., Forsyth, P.~A., Labahn, G., 2009. Valuing guaranteed
  minimum death benefit clause with partial withdrawals. Applied Mathematical
  Finance 16~(6), 451--496.

\bibitem[{Chen and Forsyth(2008)}]{ChenForsyth2008}
Chen, Z., Forsyth, P.~A., 2008. A numerical scheme for the impulse control
  formulation for pricing variable annuities with a guaranteed minimum
  withdrawal benefit (gmwb). Numerische Mathematik 109~(4), 535--569.

\bibitem[{Chen et~al.(2008)Chen, Vetzal, and Forsyth}]{ChenVeFo08}
Chen, Z., Vetzal, K., Forsyth, P.~A., 2008. The effect of modelling parameters
  on the value of {GMWB} guarantees. Insurance: Mathematics and Economics 43,
  165--173.

\bibitem[{Cramer et~al.(2007)Cramer, Matson, and Rubin}]{CramerMR2007}
Cramer, E., Matson, P., Rubin, L., 2007. Common practices relating to fasb
  statement 133, accounting for derivative instruments and hedging activities
  as it relates to variable annuities with guaranteed benefits. In: Practice
  Note. American Academy of Actuaries.

\bibitem[{Crank and Nicolson(1947)}]{crank1947}
Crank, J., Nicolson, P., 1947. A practical method for numerical evaluation of
  solutions of partial differential equations of the heat-conduction type.
  Mathematical Proceedings of the Cambridge Philosophical Society 43~(1),
  50--67.

\bibitem[{Dai et~al.(2008)Dai, Kuen~Kwok, and Zong}]{DaiKwok2008}
Dai, M., Kuen~Kwok, Y., Zong, J., 2008. Guaranteed minimum withdrawal benefit
  in variable annuities. Mathematical Finance 18~(4), 595--611.

\bibitem[{Delbaen and Schachermayer(2006)}]{DelbaenSc06}
Delbaen, F., Schachermayer, W., 2006. The Mathematics of Arbitrage. Springer.

\bibitem[{Forsyth and Vetzal(2014)}]{ForsythV2014}
Forsyth, P., Vetzal, K., 2014. An optimal stochastic control framework for
  determining the cost of hedging of variable annuities. Journal of Economic
  Dynamics and Control 44, 29--53.

\bibitem[{Fung et~al.(2014)Fung, Ignatieva, and Sherris}]{FungIgSh14}
Fung, M.~C., Ignatieva, K., Sherris, M., 2014. Systematic mortality risk: An
  analysis of guaranteed lifetime withdrawal benefits in variable annuities.
  Insurance: Mathematics and Economics 58~(1), 103--115.

\bibitem[{Hirsa(2012)}]{Hirsa}
Hirsa, A., 2012. Computational Methods in Finance. Chapman and Hall/CRC
  Financial Mathematics Series.

\bibitem[{Ho et~al.(2005)Ho, Lee, and Choi}]{lee2005}
Ho, T. S.~Y., Lee, S.~B., Choi, Y.~S., 2005. Practical considerations in
  managing variable annuities.{ }\url{doi:10.1.1.114.7023.}{ } Citeseer.

\bibitem[{Huang and Kwok(2016)}]{HuangKw16}
Huang, Y.~T., Kwok, Y.~K., 2016. Regression-based {M}onte {C}arlo methods for
  stochastic control models: variable annuities with lifelong guarantees.
  Quantitative Finance 16~(6), 905--928.

\bibitem[{Hyndman and Wenger(2014)}]{HyndmanWe14}
Hyndman, C.~B., Wenger, M., 2014. Valuation perspectives and decompositions for
  variable annuities. Insurance: Mathematics and Economics 55, 283--290.

\bibitem[{Kalberer and Ravindran(2009)}]{Kalberer2009}
Kalberer, T., Ravindran, K., 2009. Variable Annuities. Risk Books.

\bibitem[{Kling et~al.(2011)Kling, F., and Rub}]{KlingRuRu11}
Kling, A., F., R., Rub, J., 2011. The impact of stochastic volatility on
  pricing, hedging, and hedge efficiency of withdrawal benefit guarantees in
  variable annuities. ASTIN Bulletin 41~(2), 511--545.

\bibitem[{Ledlie et~al.(2008)Ledlie, Corry, Finkelstein, Ritchie, Su, and
  Wilson}]{Ledlie2008}
Ledlie, M.~C., Corry, D.~P., Finkelstein, G.~S., Ritchie, A.~J., Su, K.,
  Wilson, D. C.~E., 007 2008. Variable annuities. British Actuarial Journal
  14~(2), 327--389.

\bibitem[{Luo and Shevchenko(2015{\natexlab{a}})}]{LuoShev2015a}
Luo, X., Shevchenko, P.~V., 2015{\natexlab{a}}. Fast numerical method for
  pricing of variable annuities with guaranteed minimum withdrawal benefit
  under optimal withdrawal strategy. International Journal of Financial
  Engineering 02~(03), 1550024:1--1550024:26.

\bibitem[{Luo and Shevchenko(2015{\natexlab{b}})}]{LuoShev2015b}
Luo, X., Shevchenko, P.~V., 2015{\natexlab{b}}. Valuation of variable annuities
  with guaranteed minimum withdrawal and death benefits via stochastic control
  optimization. Insurance: Mathematics and Economics 62, 5--15.

\bibitem[{Milevsky and Salisbury(2006)}]{MilSal2006}
Milevsky, M.~A., Salisbury, T.~S., 2006. Financial valuation of guaranteed
  minimum withdrawal benefits. Insurance: Mathematics and Economics 38~(1),
  21--38.

\bibitem[{Moenig and Bauer(2015)}]{MoenigB2015}
Moenig, T., Bauer, D., 2015. Revisiting the risk-neutral approach to optimal
  policyholder behavior: A study of withdrawal guarantees in variable
  annuities. Review of Finance 20~(2), 759--794.

\bibitem[{Platen and Heath(2006)}]{PlatenHe06}
Platen, E., Heath, D., 2006. A Benchmark Approach to Quantitative Finance.
  Springer.

\bibitem[{Shevchenko and Luo(2016{\natexlab{a}})}]{ShevLuo2016}
Shevchenko, P.~V., Luo, X., 2016{\natexlab{a}}. A unified pricing of variable
  annuity guarantees under the optimal stochastic control framework. Risks
  4~(3), 22:1--22:31.

\bibitem[{Shevchenko and Luo(2016{\natexlab{b}})}]{LuoShev2016}
Shevchenko, P.~V., Luo, X., 2016{\natexlab{b}}. {Valuation of Variable
  Annuities with Guaranteed Minimum Withdrawal Benefit Under Stochastic
  Interest Rate}. ArXiv:1602.03238.

\bibitem[{Sun et~al.(2016)Sun, Kelkar, Dai, and Huang}]{Sunetal16}
Sun, P., Kelkar, R., Dai, J., Huang, V., 2016. How effective is variable
  annuity guarantee hedging?{ }Milliman Research Report:
  \url{http://au.milliman.com/insight/2016/How-effective-is-variable-annuity-guarantee-hedging}.

\end{thebibliography}

\end{document}